\def\year{2023}\relax
\documentclass[letterpaper]{article} 
\usepackage{aaai23}  
\usepackage{times}  
\usepackage{helvet} 
\usepackage{courier}  
\usepackage[hyphens]{url}  
\usepackage{graphicx} 
\usepackage{natbib}  
\usepackage{caption} 
\usepackage{subfigure}
\usepackage{booktabs}
\usepackage[table,xcdraw]{xcolor} 
\usepackage{siunitx}
\usepackage{dcolumn}

\title{Bridging Nations: \\Quantifying the Role of Multilinguals in Communication on Social Media}

\author {
    Julia Mendelsohn,\textsuperscript{\rm 1}
    Sayan Ghosh,\textsuperscript{\rm 2}
    David Jurgens,\textsuperscript{\rm 1}
    Ceren Budak\textsuperscript{\rm 1}
}

\affiliations {
    \textsuperscript{\rm 1} University of Michigan, School of Information\\
    \textsuperscript{\rm 2} University of Southern California, Department of Computer Science\\
    juliame@umich.edu, ghoshsay@usc.edu, jurgens@umich.edu, cbudak@umich.edu
}

\begin{document}
\maketitle

\begin{abstract}
  Social media enables the rapid spread of many kinds of information, from memes to social movements. However, little is known about how information crosses linguistic boundaries. We apply causal inference techniques on the European Twitter network to quantify multilingual users' structural role and communication influence in cross-lingual information exchange. Overall, multilinguals play an essential role; posting in multiple languages increases betweenness centrality by 13\%, and having a multilingual network neighbor increases monolinguals' odds of sharing domains and hashtags from another language 16-fold and 4-fold, respectively. We further show that multilinguals have a greater impact on diffusing information less accessible to their monolingual compatriots, such as information from far-away countries and content about regional politics, nascent social movements, and job opportunities. By highlighting information exchange across borders, this work sheds light on a crucial component of how information and ideas spread around the world. 
\end{abstract}

\section{Introduction}

Social media facilitates worldwide diffusion of many forms of content, such as pop culture memes, protest movements, and misinformation \citep{bruns2013arab,nissenbaum2018meme,bridgman2021infodemic}. However, most connections on social media are formed between people who share a common nationality or language \citep{ugander2011anatomy,takhteyev2012geography}. How can information spread around the world when relatively few ties cross geographic and linguistic boundaries? Multilingual users are believed to be an important piece of this puzzle, but understanding how they act as brokers in information flow across language communities remains underexplored \citep{hong2011language}. We carry out a set of causal inference studies to quantify how multilingual users influence cross-lingual information exchange across Europe on Twitter and show that the role of multilingual users varies depending on the relationship between countries and the topic of content shared.

\begin{figure}[ht]
    \centering
    \includegraphics[width=.75\columnwidth]{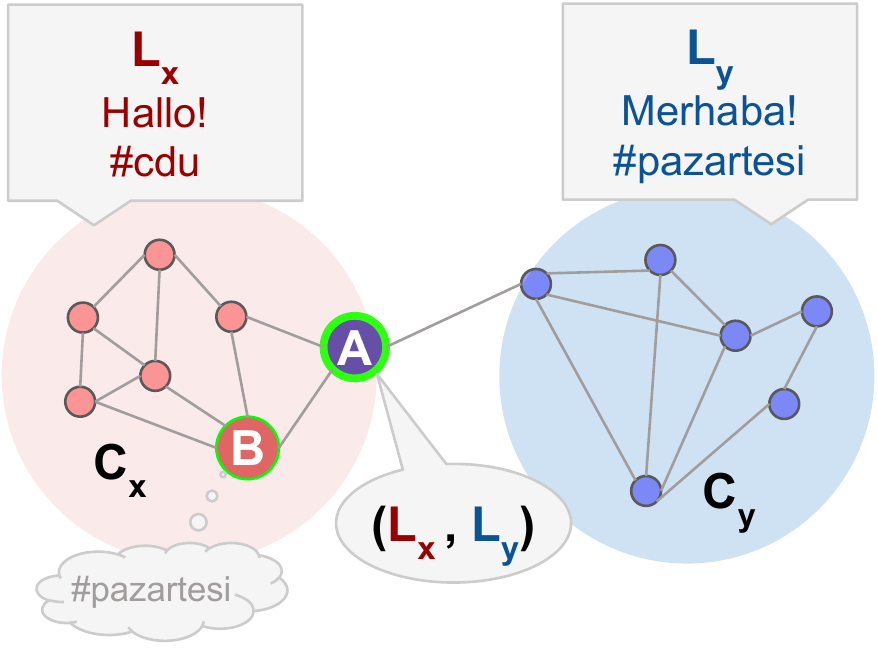}
    \caption{Consider networks of users from pairs of countries $(C_x,C_y)$ (e.g. Germany and Turkey) with dominant languages $L_x$ and $L_y$ (e.g. German and Turkish). Here, multilingual user A posts in both languages. We use causal inference to quantify (i.) the structural role and (ii.) the communication influence of A in cross-lingual information exchange. 
    }
    \label{fig:network_cartoon}
\end{figure}

We identify multilinguals based on the languages of their authored posts and avoid making claims about users' offline language competencies, which is an ideologically and theoretically fraught notion \citep{cheng2021problematizing}. Using this operationalization of multilingualism, prior work has found that multilingual users tend to have ties (e.g., following links) to multiple distinct language communities, suggesting that they act as bridges in online social networks \citep{eleta2012bridging,hale2014global}. We expand upon earlier population-level descriptive work by using causal inference techniques, namely propensity score stratification, to more robustly isolate the effects of individual users' multilingual behaviors on several outcomes. This approach is motivated by recent work showing propensity score stratification aligns with experimental results when measuring peer effects in link sharing behavior on social media \cite{eckles2020}. We carry out two analyses using this framework. First, we measure the \textbf{structural role} of multilinguals. Here, we quantify the extent to which these users act as bridges using betweenness centrality \citep{freeman1978centrality}. We then measure multilingual users' \textbf{communication influence}. To do so, we ask how having a multilingual contact impacts the content, specifically URL domains and hashtags, that monolingual users share from other languages.

By analyzing information spread across country pairs with different dominant languages, we show how multilinguals' influence varies based on relationships between countries. \citet{takhteyev2012geography} suggest that offline country relationships (e.g., economic agreements or migration patterns) impact transnational tie frequency online. Extending this conjecture, we hypothesize that the role of multilinguals is not uniform across all country pairs. Specifically, we expect multilinguals to have a bigger impact on country pairs that are more geographically distant or have weaker bonds, in which case these users would serve as gatekeepers of otherwise inaccessible information. 

We additionally consider actual content, which affects the rate and shape of information diffusion \citep{romero2011differences,tsur2012s}. However, this cannot be captured by analyses of network structure alone. Our measures of multilinguals' communication influence overcome this limitation by enabling us to compare effects across content topics. We again hypothesize multilingual users play a larger role in spreading content otherwise inaccessible to an international audience. For example, we expect multilinguals to have a bigger influence on spreading hashtags about local or national politics compared to widespread entertainment sensations.

Our contributions are as follows: in large-scale causal inference studies of
European Twitter, we show that multilingually-posting users play a vital structural role and communication influence on information diffusion across languages. We compare how multilinguals' influence varies based on relationships between countries and find they have a greater effect among country pairs that are more geographically distant, especially for Western Europeans who post in Eastern European languages. We further measure how the effect of multilinguals is driven by content topic and find that they have the largest influence in spreading content about politics, developing health-related social movements, and job opportunities.

Understanding how multilinguals affect information diffusion has immense consequences for online platforms. For example, platforms may focus efforts on empowering multilingual users to spread information that can support knowledge-sharing, collaboration, crisis response, social progress, or other beneficial outcomes \citep{eleta2014multilingual}. On the other hand, they may want to discourage multilinguals from sharing dangerous content such as misinformation or abuse.

\section{Related Work}

We draw upon prior work on multilinguals' online behavior and information diffusion across social networks.

\subsection{Social networks and information diffusion}

Online social networks tend to have relatively few ties connecting distinct national and linguistic communities, leading to structural holes \citep{burt2004structural,ugander2011anatomy,hale2012net}. Spreading novel information across these communities thus relies on bridges that span structural holes \citep{easley2010networks}. Information spreads the most quickly across long-range bridges, where the intermediary node greatly reduces the shortest path between two other nodes \citep{kossinets2008structure}. Bridges in a social network play a brokering role that adds to their social capital \citep{burt2004structural}. 
We posit that multilinguals play an important role in cross-lingual information exchange because they serve as bridges between language communities. 

Prior work measures how people influence others to propagate information \citep{guille2013information}. For example, people are more likely to share information their friends share, and overall, weak ties (e.g., acquaintances) are more responsible for spreading novel information than strong ties (e.g., close friends) \citep{bakshy2012role}. Although multilinguals are not necessarily weak ties, they similarly can act as bridges between communities. Methodologically, we draw from \citet{eckles2020}, who use causal inference techniques, namely propensity score stratification, to measure peer influence on link-sharing behavior and show that their observational estimates are consistent with experimental results.

Understanding information diffusion and influential brokers impacts research across disciplines, including the development of activism and protest movements \citep{gonzalez2011dynamics,park2015comparing,lee2020global}, spread of misinformation \citep{bridgman2021infodemic}, product adoption \citep{talukdar2002investigating}, and online abuse \citep{sporlein2021ethnic}.

\subsection{The bridging role of multilinguals}

Prior work has analyzed connections within and across language communities online \citep{hale2012net,hale2014multilinguals,samoilenko2016linguistic}. \citet{hale2014multilinguals} argues that multilingual editors on Wikipedia are important for sharing knowledge across language communities and facilitate access to more diverse knowledge; they are more active than their monolingual counterparts overall and often write the same article across multiple language versions. However, \citet{kim2016understanding} suggests that language remains a barrier because multilinguals are less engaged and less likely to edit complex content in a non-primary language. While online bloggers primarily link to content within the same language, cross-lingual links do exist, and some bloggers explicitly seek to connect distinct language communities \citep{zuckerman2008meet,hale2012net}. Authors of blogs that bridge language communities often tend to be multilingual migrants or language learners \citep{herring2007language}. On Twitter, multilinguals disseminate information across different public spheres during events such as the Arab Spring \citep{bruns2013arab}. 

Earlier work has also examined the structural role of multilinguals. \citet{kim2014sociolinguistic} count edges between monolingual and multilingual ``lingua" groups within multilingual regions on Twitter. They find that monolingual groups tend to be connected via multilingual groups, suggesting that multilingual users are bridges between different language communities. \citet{hale2014global} similarly argues that multilinguals collectively play an important bridging role in the Twitter network. When removing all multilingual nodes, the largest connected component becomes smaller and the number of small components increases, and these changes are significantly larger than if the same number of randomly-selected monolinguals were removed \citep{hale2014global}. \citet{eleta2012bridging,eleta2014multilingual} characterize multilingual users' ego-networks as gatekeepers (language communities connected by only a few users) and language bridges (tightly-connected language groups). The authors suggest that gatekeepers are essential individuals for spreading information across linguistic, national, and cultural boundaries. 

Other work has measured multilinguals' participation in information cascades (resharing chains) on Twitter. \citet{jin2017detection} find that multilingual behaviors of the original poster and their followers are predictive of information cascades crossing languages. \citet{chen2021burden} compare monolingual and multilingual users in cascade trees of COVID-19 information, develop measures that capture users' bridging roles based on how much they propagate information, and find that multilinguals have a bigger bridging role than monolinguals in nearly two-thirds of information cascades.

We build upon prior work in several important ways. While structural analyses suggest that multilinguals are important for cross-lingual information exchange, our causal inference studies quantify the effect of multilingual behaviors on both structural and content-sharing outcomes while accounting for possible confounds such as posting frequency and overall popularity. We further extend the evidence that online multilingual behaviors and connections vary across countries by identifying systematic variation in the effect of multilinguality across country pairs \citep{kim2014sociolinguistic}.

\begin{figure*}[htp]
  \centering
  \includegraphics[width=\linewidth]{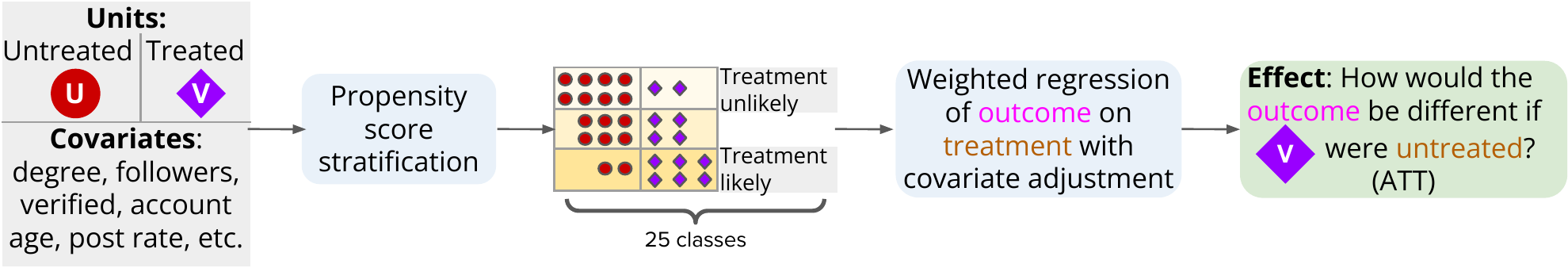}
  \caption{General design for causal inference studies \citep{ho2007matching,stuart2010matching}. We first estimate the propensity of each unit to receive treatment and use these scores to divide the sample into 25 strata. We then compare treated and untreated units within each stratum through a weighted regression to estimate the causal effect of the treatment.}
  \label{fig:study-design}
\end{figure*}

\section{Data}

We construct an undirected network with Decahose data from 2012-2020, where edges are mutual ``mentions" between users. We consider all pairs of European countries $(C_x,C_y)$ with dominant languages $L_x$ and $L_y$, and select countries based on geography and Eurovision participation, a marker of cultural affiliation. Using location inference \cite{compton2014geotagging}, we extract network subsets for all $(C_x,C_y)$ pairs where nodes are users from $C_x$ or $C_y$. We use tweets from 2018-2020 for multilingualism and content measures due to limited text availability.

\paragraph{Definitions} Each pair of countries $(C_x,C_y)$ is a \textit{multilingual country pair (MCP)} with dominant languages $L_x$ and $L_y$. Users who post only in $L_x$ are \textit{$L_x$ monolinguals} and users who post in both $L_x$ and $L_y$ are \textit{$(L_x,L_y)$ multilinguals}. Within each $(C_x,C_y)$ network we separately measure the role of $(L_x,L_y)$ multilinguals located in $C_x$ and $C_y$, which we refer to as \textit{loci}. In Figure \ref{fig:network_cartoon}, \textit{(Germany, Turkey)} is an MCP, and we measure the role of \textit{(German, Turkish)} multilinguals within each locus \textit{Germany} and \textit{Turkey}.

\paragraph{Language identification}

Twitter posts present a challenge to automated language identification (LangID) due to their short length, informal style, and lack of ground truth labels \citep{graham2014world,williams2017twitter}. We use Twitter's built-in language detector because it is computationally efficient for a massive dataset, requires few additional resources, and is trained on in-domain data.\footnote{https://blog.twitter.com/engineering/en\_us/a/2015/\\evaluating-language-identification-performance}

We validate our decision with a comparison to 5 popular LangID packages: \texttt{langdetect}\footnote{https://github.com/shuyo/language-detection}, \texttt{langid.py} \citep{lui2012langid}, and \texttt{CLD2}\footnote{https://github.com/CLD2Owners/cld2} use probabilistic models, while \texttt{fastText} \citep{joulin2016fasttext} and \texttt{CLD3}\footnote{https://github.com/google/cld3} use neural networks. As in prior LangID evaluations, we randomly sample 1K tweets from 32 countries written in that country's dominant language, as labeled by Twitter \citep{graham2014world,lamprinidis2021universal}. Like \citet{graham2014world}, we calculate intercoder agreement between all pairs of models. Table \ref{tab:langid} (Supplemental Material) shows that Twitter's LangID has a high agreement with other models, even at higher rates than they agree with each other.

Like most LangID models, Twitter's LangID has relatively low coverage of the world's languages and may struggle to distinguish similar languages \citep{lui2014accurate,williams2017twitter}. We mitigate these limitations by selecting multilinguals and MCPs so that our analyses are not overly sensitive to individual prediction errors.

\paragraph{Identifying multilingual users}

Following \citet{eleta2014multilingual}, an individual uses language $L$ if at least 10\% of their posts containing original content are written in $L$ (i.e., excluding retweets but including quote tweets and replies). A user is \textit{monolingual} if one language passes the 10\% threshold, and \textit{multilingual} in the $(C_x,C_y)$ network if both $L_x$ and $L_y$ pass this threshold. Following \citet{kim2014sociolinguistic}, we collect language information for all users with at least five posts in the Decahose. We additionally exclude users if over 20\% of their tweets are in unidentified languages because our calculations for language use may be less accurate. We set these thresholds so that estimates of users' multilingualism are more reliable and robust to language prediction errors of individual tweets. We emphasize that our operationalization of multilingualism is based solely on users' expression on Twitter; we do not make claims about language knowledge or offline behavior.

\paragraph{Selecting network subsets for analysis}

From an initial set of 50 European countries, we exclude 18 after three filtering steps. First, we remove six micro-states with area under 500km$^2$ because \citet{compton2014geotagging}'s reported errors suggest that location inference is less reliable within such small areas. Second, we restrict our analysis to countries with a single official or dominant language (i.e., used by at least 70\% of the population), removing eight more countries.\footnote{Data for the geographic area and population-level language usage is from the CIA World Factbook.} This step is necessary for our problem formulation, which focuses on information spread across borders. The study of users within highly-multilingual countries is beyond the scope of this paper. Third, we calculate the distribution of tweets authored by users from each country and exclude four countries where the majority of tweets' languages are unidentified or no tweets are identified as written in the country's dominant language. See Supplemental Material Table \ref{tab:countries} for the list of included and filtered countries.

We include an MCP $(C_x,C_y)$ if: (1) $C_x$ and $C_y$ have different dominant languages, and (2) the $(C_x,C_y)$ network is sufficiently large to rigorously estimate causal effects. Specifically, an MCP network is included if it has at least 100 $L_x$ and $L_y$ monolinguals, 20 $(L_x,L_y)$ multilinguals, and 100 users with at least 1 $(L_x,L_y)$ multilingual neighbor.\footnote{While we deem such thresholds necessary for precise causal effect estimation, we acknowledge the arbitrariness of these numbers as a limitation of our study.} See Supplemental Material Table \ref{tab:mcp} for included MCPs.

\paragraph{Descriptive statistics}
Our dataset contains information for 12.6M users from 32 countries. 8.0M users pass the thresholds for posting activity and language identifiability and are considered for analysis. Of these users, 1.1 million (13.6\%) tweet in both their country's dominant language and another European language. Norway has the largest percentage of bilingual users (46.9\%), while the United Kingdom has the least (3.3\%). 232 MCP network subsets were selected based on our inclusion criteria, covering 30 European countries and 7.7 million unique users (Georgia and Moldova were considered, but no MCP network containing these countries was included). MCP networks range in size from 4.7K nodes \textit{(Latvia, Lithuania)} to 3.6M nodes \textit{(United Kingdom, Turkey)}, with a median size of 319K nodes.

\section{Problem Formulation}

Adopting the Neyman-Rubin framework of potential outcomes \citep{holland1986statistics}, we isolate the effect of multilingualism by addressing the counterfactual: how would a multilingual user $u$'s influence be different if $u$ were monolingual? Each study's details vary (Table \ref{tab:formulation}), but all have the same idea: we define a set of \textit{units} (users), some of whom receive a \textit{treatment} indicating multilingual behavior, and we estimate an \textit{outcome} variable related to influence on information exchange across languages.

\begin{table}[!htbp]
\caption{Overview of problem formulation for causal inference studies. Each study focuses on the effect of multilingualism on cross-lingual information exchange, but defines different samples, treatments, and outcomes.}
\label{tab:formulation}
\resizebox{\linewidth}{!}{
\begin{tabular}{@{}cccc@{}}
\toprule
 & Structural Role & \multicolumn{2}{c}{Communication Influence} \\ \midrule
 & Study 1 & Study 2 & Study 3 \\ \midrule
Units & \begin{tabular}[c]{@{}c@{}}$C_x$ users\\ posting in $L_x$\end{tabular} & \begin{tabular}[c]{@{}c@{}}Monolinguals\\  in $L_x$ from $C_x$\end{tabular} & \begin{tabular}[c]{@{}c@{}}Monolinguals\\  in $L_x$ from $C_x$\end{tabular} \\ \midrule
Treatment & \begin{tabular}[c]{@{}c@{}}Posting \\multilingually \\ in $(L_x,L_y)$\end{tabular} & \begin{tabular}[c]{@{}c@{}}Having $\geq1$ \\ $(L_x, L_y)$ multilingual \\ neighbor\end{tabular} & \begin{tabular}[c]{@{}c@{}}Having $\geq1$ \\ $(L_x,L_y)$ multilingual \\ neighbor\end{tabular} \\ \midrule
Outcome & \begin{tabular}[c]{@{}c@{}}Betweenness \\ centrality\end{tabular} & \begin{tabular}[c]{@{}c@{}}Sharing domain \\ from $L_y$\end{tabular} & \begin{tabular}[c]{@{}c@{}}Sharing hashtag \\ from $L_y$\end{tabular} \\ \bottomrule
\end{tabular}}
\end{table}

\subsection{Measures of influence}

Are multilinguals well-positioned in networks to spread information? In Study 1, we quantify  their structural role by estimating the effect of multilingual posting on betweenness centrality. A measure of the proportion of shortest paths that must go through an intermediary node, betweenness centrality quantifies the extent to which the intermediary is an information broker \citep{freeman1978centrality}. Studies 2 and 3 focus on multilinguals' influence on content flows across languages. If an $L_x$ monolingual has a multilingual neighbor in the $(C_x,C_y)$ network, how much more likely are they to share $L_y$ content? We consider two forms of content: URL domains and hashtags \citep{hong2011language}. 

\paragraph{Betweenness centrality (Study 1)}
For each $(C_x,C_y)$ MCP, we consider a sample of users from locus $C_x$ who post in $L_x$. The treatment is if a user is multilingual in $(L_x,L_y)$, and the outcome is the user's betweenness centrality in the $(C_x,C_y)$ network, scaled by $10^6$ and log-transformed.

\paragraph{Domain sharing (Study 2)}

We first identify URL domains associated with each language. We filter out the 200 overall most-frequent domains. Like stop word removal \citep{jurafsky2000nlp}, excluded domains are largely uninformative for measuring content sharing across languages (e.g. \textit{instagram.com}, \textit{unfollowspy.com}). To account for sampling variability, we exclude domains that appear fewer than ten times \citep{monroe2008fightin}. We then consider ``domains associated with $L_x$" to be the 100 domains most overrepresented in $L_x$ tweets (excluding retweets) relative to all European tweets based on the weighted log odds ratio with an informative Dirichlet prior \citep{monroe2008fightin}. For each MCP $(C_x,C_y)$ and locus $C_x$, the sample is the set of $L_x$ monolinguals from $C_x$. A user $u$ is treated if they have at least one $(L_x,L_y)$ multilingual neighbor, and the outcome is whether any of $u$'s Decahose tweets contain at least one domain associated with $L_y$, the language $u$ does \textit{not} use. Note that we exclude retweets for language-based measures, such as identifying multilinguals and language-specific domains/hashtags, but include retweets for information-sharing measures because they are an important component of information diffusion on Twitter.

\paragraph{Hashtag sharing (Study 3)}

We similarly identify hashtags associated with each language. Unlike domains, language-associated hashtags can change rapidly because they may refer to short-term events, such as elections, protests, or upcoming TV shows. We thus separate the Decahose data into 60 14-day intervals, and use the log-odds ratio with informative Dirichlet priors to identify $H_x^i$, the set of 100 hashtags most associated with $L_x$ in interval $i$, after again filtering out 200 most-frequent hashtags (e.g. \textit{\#blog}, \textit{\#radio}), those occurring fewer than ten times in $i$, and excluding retweets. A user $u$ ``shares" a hashtag from $L_x$ if any of $u$'s Decahose tweets contain at least one hashtag $h \in H_x^i$ during $i$ or the subsequent period $i+1$ (including retweets). Resulting hashtags reflect entertainment, sports, politics, and everyday life; see Supplemental Material Table \ref{tab:examples} for examples of language-specific domains and hashtags. 

\subsection{Causal inference setup} 

We use propensity score stratification to estimate the causal effects of multilingual treatment variables on information diffusion outcomes \citep{rosenbaum1983central}. Our procedure, shown in Figure \ref{fig:study-design}, closely follows the guidelines provided by \citet{stuart2010matching} and MatchIt \citep{ho2007matching}.  

We first fit a logistic regression model to calculate propensity scores, which represent the probability of receiving treatment as a function of the specified covariates \citep{rosenbaum1983central}. Covariates are shared for all studies and capture aspects of users' Twitter behavior that may affect our outcomes \citep{stuart2010matching}. These include verified status, network degree, follower and following counts, account age, number of tweets in the Decahose sample, ``favorites" count, and post rate, all log-scaled. Covariates for each user are based on their most recent tweet in our sample. 

We then conduct propensity score stratification where units are separated into 25 strata based on propensity scores and verify sufficient balance for all covariates with absolute standardized mean difference less than 0.1 \citep{ho2007matching}. Although five strata have commonly been used in practice, more strata can yield less biased estimates for larger samples sizes \citep{eckles2020}. Propensity score estimation, stratification, and balance checks are performed using the MatchIt R package \citep{ho2007matching}.

We compare treated and untreated units within each stratum by fitting a regression of each outcome on treatment status, weighted by the matching weights. In particular, we estimate the average treatment effect on the treated (ATT); this estimates how the outcome among treated units is different than in the counterfactual scenario where they are not treated \citep{stuart2010matching}. For all studies, we include covariate adjustment to control for direct effects that pre-treatment covariates may have on the outcome. 

Because each study defines units, treatment, and outcomes differently, the specific details of ATT estimation vary. In Study 1, we fit a linear regression after propensity score stratification to estimate the difference in (scaled and log-transformed) betweenness centrality. Here, multilingualism increases betweenness centrality if ATT $>$0. In Studies 2 and 3, the outcomes are binary so we use logistic regression to estimate ATT as an odds ratio, a measure used in prior work to compare domain and hashtag sharing behaviors across languages \citep{hong2011language}. For some treated user $u$ who is monolingual in $L_x$, the ATT estimates the odds that $u$ shares a domain (hashtag) associated with $L_y$ divided by the odds that $u$ would share a domain (hashtag) associated with $L_y$ in the counterfactual scenario where $u$ is not treated. ATT $>$ 1 indicates that having a multilingual $(L_x,L_y)$ contact increases an $L_x$ monolingual's likelihood of sharing domains (hashtags) from $L_y$. 

In each study, we estimate the ATT on aggregate data from both loci in all MCP networks in order to get a single causal estimate of the role of multilinguals in cross-lingual information exchange. We additionally estimate separate ATT values for units from each locus in each MCP network, and analyze the variation across locus-specific causal effects later. To ensure sufficient data across strata, we only estimate locus-specific causal effects for locus $C_x$ in MCP $(C_x,C_y)$ if there are at least 100 treated units. Furthermore, we only include locus-specific causal effects for Studies 2 and 3 if at least 100 units in the combined treated and untreated sample (all $L_x$ monolinguals) have an outcome of 1 (share at least one domain or hashtag from $L_y$).

\section{Overall effects of multilingual behavior}

\begin{table}[]
\caption{Summary of causal effect of posting multilingually (Study 1) or having a multilingual friend (Studies 2 and 3) for users in each country (locus) across all multilingual country pairs (MCPs).  Despite considerable variation across loci, each treatment has positive effects on information diffusion-related outcomes far more than negative effects. We estimate are average treated effect on the treated (ATT), and significance is determined at $p < 0.05$ with robust standard error estimation. }
\label{tab:overall_results_proportions}
\resizebox{\columnwidth}{!}{%
\begin{tabular}{@{}cccc@{}} \toprule
 & \multicolumn{3}{c}{Outcome} \\ \midrule
 & \begin{tabular}[c]{@{}c@{}}Betweenness \\ Centrality\\ (Study 1)\end{tabular} & \begin{tabular}[c]{@{}c@{}}Domain \\ Sharing\\ (Study 2)\end{tabular} & \begin{tabular}[c]{@{}c@{}}Hashtag \\ Sharing\\ (Study 3)\end{tabular} \\ \midrule
\# Eligible MCPs & 214 & 158 & 199 \\
\# Eligible Loci & 317 & 205 & 284 \\
\% Loci w/ sig. pos ATT & 46.37\% & 56.10\% & 50.00\% \\
\% Loci w/ no sig. ATT & 51.42\% & 40.49\% & 46.48\% \\
\% Loci w/ sig. neg ATT & 2.21\% & 3.41\% & 3.52\% \\ \bottomrule
\end{tabular}%
}
\end{table}

All 3 studies support our hypothesis that multilinguals play an important role in cross-lingual information exchange. 

\paragraph{Betweenness centrality (Study 1)}
Multilingual $(L_x,L_y)$ users have higher betweenness centrality in the $(C_x,C_y)$ network than their monolingual peers, suggesting that these users serve as local bridges and thus are well-positioned to spread novel information across the network \cite{freeman1978centrality,granovetter1973strength}. The overall ATT is 0.034 ($p < 0.0001$, with robust standard error estimation). In other words, multilingual posting increases the outcome of log-transformed betweenness by 0.034, which corresponds to a 13.5\% increase in betweenness centrality. 

\paragraph{URL domain sharing (Study 2)}
Having a multilingual $(L_x,L_y)$ friend increases the odds of a monolingual $L_x$ user sharing a domain from $L_y$ by a factor of 15.57 ($p < 0.0001$). For interpretability, we also estimate the average marginal effect with a marginal effects model on the logistic regression used for estimating the ATT as an odds ratio. We find that having a $(L_x,L_y)$ friend increases an $L_x$ monolingual's probability of sharing an $L_y$ domain by 20.0\%.

\paragraph{Hashtag sharing (Study 3)}
Having a multilingual $(L_x,L_y$) friend significantly increases the odds of an $L_x$ monolingual sharing a hashtag from $L_y$ by a factor of 3.98 ($p < 0.0001$). Through estimating the average marginal effect, we find that this corresponds to an increase in the probability of sharing an $L_y$ hashtag by 32.7\%. While the odds ratio for hashtag sharing is lower than for domains, the probability increase is greater because extremely low domain-sharing probabilities result in inflated odds ratios.

Table \ref{tab:overall_results_proportions} summarizes locus-specific causal effect estimates. Due to minimum inclusion criteria, we only estimate ATTs for a subset of all 464 loci corresponding to 232 MCPs. All three treatments increase information-sharing outcomes in about half of the loci, and decreases outcomes in under 4\% percent of loci. Locus-specific estimates reinforce that multilinguals facilitate information exchange across language boundaries. However, the distribution of positive, negative, and insignificant effects across loci as shown in Table \ref{tab:overall_results_proportions} suggests wide variation across MCPs and loci.

\begin{figure*}[htbp!]
   \centering
    \includegraphics[width=.5\linewidth]{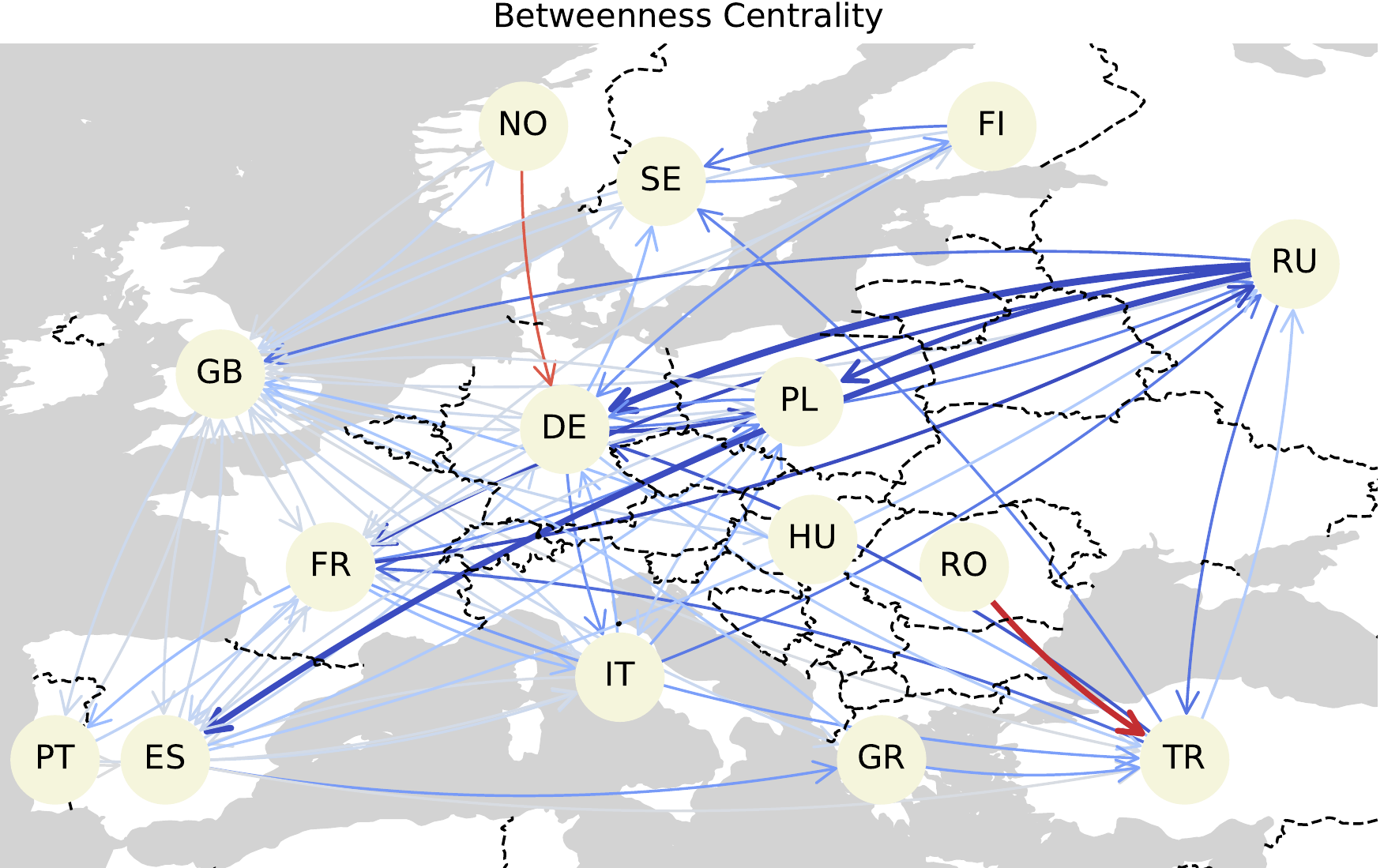}\hfill
    \includegraphics[width=.5\linewidth]{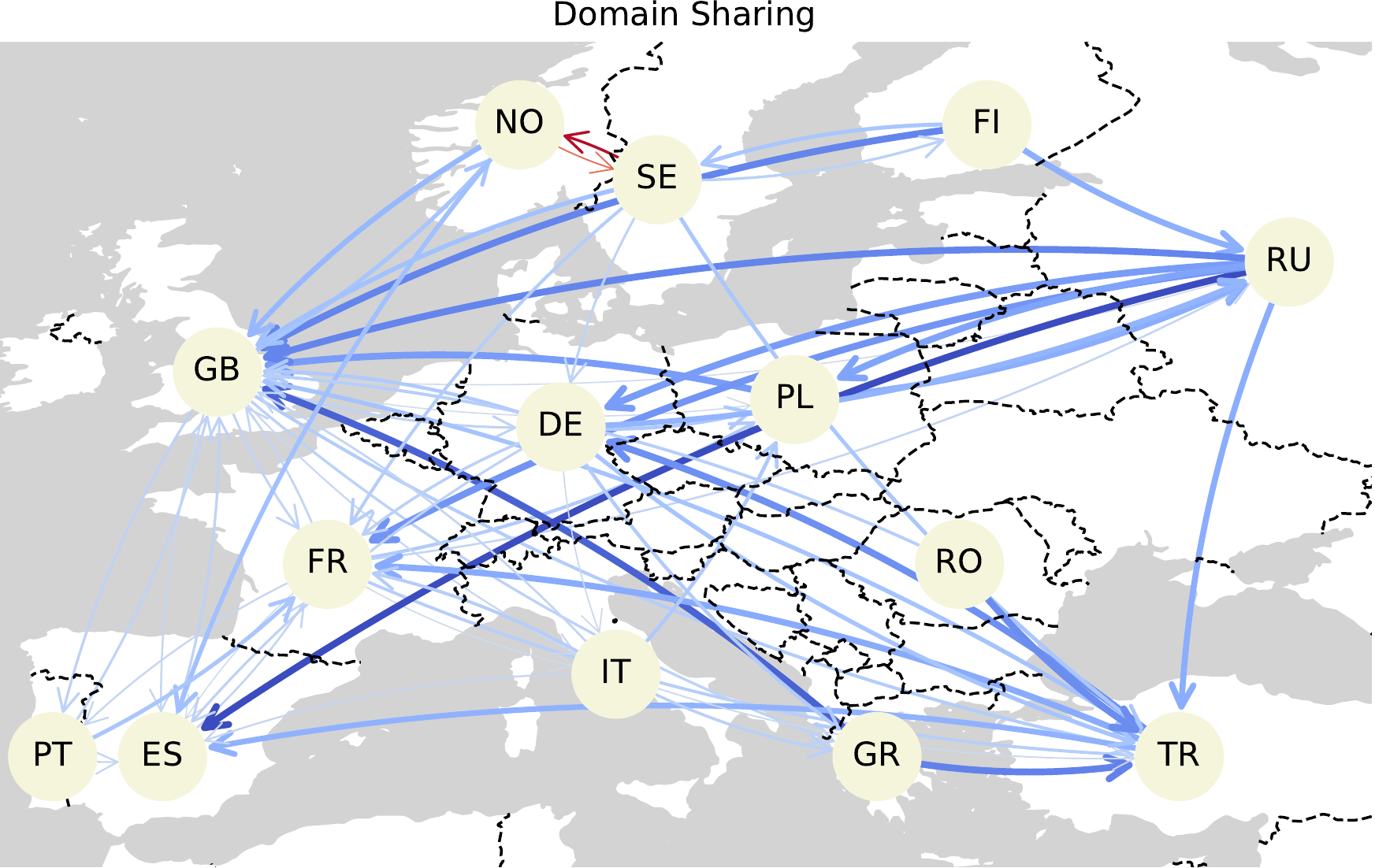}\hfill
    \caption{Maps of causal effects (ATTs) of multilingual treatments on information diffusion outcomes. The left map shows the ATTs of multilingual posting on betweenness centrality (Study 1), where each edge indicates ATT among users from the \textit{destination} node who post in both the source and destination nodes' dominant languages. The right map shows the ATTs of having a multilingual friend on cross-lingual domain sharing (Study 2), where each edge indicates ATT among monolingual users from the \textit{destination} node on sharing URL domains associated with the \textit{source} node's dominant language. Negative effects are shown with red arrows and positive are shown with blue arrows. The magnitude of the causal effect is indicated by arrow shading and width. Only statistically significant estimates ($p<0.05$) with robust standard error estimation are shown.}
    \label{fig:maps}
\end{figure*}

\section{Variation across country pairs}
All three studies show that multilingual behaviors increase cross-lingual information exchange. However, locus-specific causal estimates indicate substantial heterogeneity in the magnitude of their effects across MCPs and loci. 
We conduct regression analyses to characterize how geographic, demographic, economic, political, and linguistic relationships between country pairs correlate with the effects of multilingual behaviors on European Twitter.

\subsection{Regression Setup}

We fit linear regression models to characterize how the causal effects of multilinguals vary across countries. For a given MCP $(C_x,C_y)$ and locus $C_x$, we define 3 dependent variables: the estimated causal effects (ATT) of multilingual behavior in each of the 3 described studies. Independent variables capture relationships between $C_x$ and $C_y$. 

\textbf{Demographic variables} include the population ratio of $C_x$ to $C_y$ from 2019 based on CEPII's Gravity Dataset \citep{head2014gravity,conte2021cepii}. Using 2017 World Bank migration data, we consider the fraction of $C_x$'s population born in $C_y$, $C_y$'s population born in $C_x$, and each country's population who are foreign-born.\footnote{\url{worldbank.org/en/topic/migrationremittancesdiasporaissues/brief/migration-remittances-data}}

\textbf{Geographic variables} are the distance (in km) between $C_x$ and $C_y$'s population centers and time difference: the number of hours $C_y$ is ahead (further east) of $C_x$.

\textbf{Economic variables} include i.) the ratio of $C_x$ to $C_y$'s GDP per capita, ii.) if $C_x$ and $C_y$ are in an RTA (Regional Trade Agreement, which includes the EU), and iii.)  trade flow between $C_x$ and $C_y$ averaged over both country's reports and both directions, normalized by the total population of both countries. Geographic and economic variables use 2019 data from CEPII's Gravity Dataset.

\textbf{Political variables}, specifically material conflict between $C_x$ and $C_y$, are determined by querying the GDELT event database \cite{leetaru2013gdelt}. 
These include the percent of $C_x$'s external conflict actions inflicted on $C_y$, and $C_y$'s external conflict actions inflicted on $C_x$.   

The last fixed effect is \textbf{linguistic distance} between $L_x$ and $L_y$ using Glottolog \citep{hammarstrom2021glottolog}. Inspired by \citet{samoilenko2016linguistic}'s measurement of shared language families, 
we define a 4-level measurement of linguistic distance between $L_x$ and $L_y$: i.) no relationship (e.g. Spanish and Hungarian), ii.) in the same primary family (e.g. German and Polish are Indo-European), iii.) in the same branch (e.g. English and Swedish are Germanic), and iv.) in the same sub-branch (e.g. Spanish and Italian are Romance).\footnote{We do not use graph-based measurements of distance because there is wide variation across language branches' structures due to an uneven interest by linguists across languages.}

We avoid multicollinearity issues by ensuring that all variables' variance inflation factor is under 4. We thus exclude highly-correlated variables, such as EU membership and each country's population. We weight each regression model by the number of treated units from each locus. Finally, we scale all variables by z-score to facilitate direct comparisons.

\begin{table}[!htbp] \centering 
  \caption{Geographic, demographic, economic, political, and linguistic aspects of relations between countries $C_x$ and $C_y$ and their associations with the role of multilinguals in spreading information from $C_y$'s dominant language to $C_x$. Coefficients are from weighted linear regression models where the dependent variables are the locus-specific causal effects (ATTs) of multilingual behaviors in Studies 1-3.} 
  \label{tab:regression} 
  \resizebox{\linewidth}{!}{
\begin{tabular}{@{\extracolsep{-25pt}}lD{.}{.}{-3} D{.}{.}{-3} D{.}{.}{-3} } 
\toprule
 & \multicolumn{3}{c}{\textit{Dependent variable:}} \\ 
\cline{2-4} 
\\[-1.8ex] & \multicolumn{1}{l}{Betweenness} & \multicolumn{1}{c}{Domain} & \multicolumn{1}{c}{Hashtag} \\ 
 & \multicolumn{1}{l}{Centrality} & \multicolumn{1}{c}{Sharing} & \multicolumn{1}{c}{Sharing}\\ 
\hline \\[-1.8ex] 
 Geographic distance & 0.020^{***} & 0.434 & 0.412^{***} \\ 
  Time difference & 0.015^{**} & 0.815^{***} & 0.001 \\ \midrule
  Pop. $C_x$ / $C_y$ & -0.030^{***} & -0.031 & 0.017 \\ 
    \% $C_y$ foreign-born & 0.021^{**} & -1.100^{**} & -0.058 \\ 
  \% $C_y$ pop. born in $C_x$ & 0.017^{**} & -0.044 & -0.043 \\ 
    \% $C_x$ foreign-born & -0.002 & -0.031 & 0.064 \\ 
  \% $C_x$ pop. born in $C_y$ & 0.007 & -0.197 & -0.040 \\ \midrule
  GDP per capita $C_x$ / $C_y$ & 0.038^{***} & 0.318 & 1.171^{***} \\ 
  RTA & 0.010 & -0.411 & 0.216^{*} \\ 
  Tradeflow per capita & -0.013^{*} & 0.425 & 0.109 \\ \midrule
  \% $C_x$'s conflicts vs. $C_y$ & 0.019^{**} & -0.159 & -0.032 \\ 
  \% $C_y$'s conflicts vs. $C_x$ & 0.002 & -0.257 & 0.088 \\ \midrule
  Linguistic distance & -0.027^{***} & 0.449 & 0.146^{*} \\ \midrule

Observations & \multicolumn{1}{c}{317} & \multicolumn{1}{c}{205} & \multicolumn{1}{c}{284} \\ 
R$^{2}$ & \multicolumn{1}{c}{0.266} & \multicolumn{1}{c}{0.193} & \multicolumn{1}{c}{0.448} \\ 

\bottomrule 
\textit{Note:}  & \multicolumn{3}{r}{$^{*}$p$<$0.1; $^{**}$p$<$0.05; $^{***}$p$<$0.01} \\ 
\end{tabular} }
\end{table} 

\subsection{Regression results}

\subsubsection{Geography}
Multilinguals have a larger effect on cross-lingual information exchange in MCPs where $C_x$ and $C_y$ are further away from each other (Table \ref{tab:regression}). We visualize this pattern in Figure \ref{fig:maps}, which shows the relationship between geography and effects of multilinguals on betweenness centrality and domain sharing (see Figure \ref{fig:hashtag_maps} in Supplemental Material for the hashtag sharing map). 

In Figure \ref{fig:maps}, the effect of multilinguals in MCP $(C_x,C_y)$ and locus $C_x$ is shown as a directed edge from $C_y$ to $C_x$. Only edges corresponding to significant estimates are drawn. Negative effects are red, positive effects are blue, and greater magnitude is represented with darker and thicker edges. For example, the dark blue edge from Russia to Spain in both maps indicates that Russian-Spanish multilinguals are especially important for bringing Russian information to Spain. In contrast, the faint edge from Portugal to Spain means that Portuguese-Spanish multilinguals have a smaller role in importing Portuguese information to Spain. 

We believe stronger treatment effects across longer distances are due to information accessibility. Information from faraway places is not readily accessible for monolinguals, so they may need to rely more on their multilingual friends to serve as information brokers. On the other hand, information between nearby countries such as Norway and Sweden may be more easily accessible with more channels for diffusion, possibly via more multilinguals, so users rely less on individual multilinguals to spread information. 

In addition to distance, the time diference between $C_x$ and $C_y$ significantly predicts multilinguals' effects (Table \ref{tab:regression}). Multilinguals have the largest impact when $C_y$ is 2-3 hours ahead (i.e., further east) of $C_x$ (see Figure \ref{fig:time_diff} in Supplemental Material). In other words, multilinguals have a large impact on spreading information from Eastern European languages to Western Europe. This asymmetric East-West pattern is visible in both maps of Figure \ref{fig:maps} (e.g. the many dark blue outgoing links from Russia indicate a large influence of multilingual users from other countries who post in Russian).  

Why are Western European users of Eastern European languages so influential in cross-lingual information exchange? Fully answering this question is left for future work, but we speculate that it is a consequence of historical inaccessibility to information across strict Cold War-era borders. Additionally, offline connections may explain these users' online role as bridges; Eastern migrants in Western Europe have been characterized by more transnational and circular offline networks since the early 2000s \citep{favell2008new}.

\subsubsection{Demographics} 
Multilingualism increases betweenness centrality more for users in smaller countries who post in more populous countries' languages, but this relationship is not significant for content sharing outcomes. While multilinguals in smaller countries are better positioned within MCP networks to spread novel information, they do not necessarily ``import" information from the larger to smaller country. 

When $C_y$ has a higher proportion of foreign-born residents, $(L_x,L_y)$ multilingualism has a greater impact on users' structural role in $C_x$, but not communication influence. In fact, having a multilingual friend has a negative impact on sharing a domain from $L_y$ when $C_y$ has more foreign-born residents; perhaps this is because shared links from $L_y$ feature more content with multicultural or international appeal, so people rely less on multilinguals as information brokers. We observe no significant effects of the foreign-born population of $C_x$ on either the structural role or communication influence of multilinguals.

\subsubsection{Economics and politics} 
Multilinguals' effects are associated with GDP per capita: multilinguals in $C_x$ have more influence when $C_x$ is wealthier than $C_y$ (although this is not significant for domain sharing). However, government-level relationships including national trade agreements, tradeflows, and political conflict are not significant predictors. 

\subsubsection{Linguistic similarity}
We observe mixed results for linguistic distance. Our information accessibility hypothesis suggests $(L_x,L_y)$ multilinguals would play a larger role when $L_x$ and $L_y$ are very different because their monolingual peers would rely more heavily on them. This pattern appears to hold for communication influence outcomes but is not significant. Contrary to our hypothesis, multilinguals' structural role is amplified when $L_x$ and $L_y$ are closely related. Though we do not have a clear explanation for this pattern, it may be driven by deeper differences in the structures of MCP networks connecting countries with similar dominant languages \citep{herring2007language}. 

\section{Variation across topics}
Different aspects of content, such as topic, framing, sentiment, and subjectivity, could amplify or hinder multilinguals' influence. 
Therefore, here we extend Study 3 to investigate how the role of multilinguals depends on the topic of shared hashtags using unsupervised topic modeling \citep{kim2014sociolinguistic,jin2017detection}. Specifically, for each MCP $(C_x,C_y)$ we measure how having a multilingual $(L_x,L_y)$ friend affects the odds of an $L_x$ monolingual from $C_x$ sharing a \textit{topic-specific hashtag} from $C_y$.

\begin{table}[!htbp]
\caption{Fifteen topics of interest with example hashtags}
\label{tab:topic_examples}
\resizebox{\linewidth}{!}{
\begin{tabular}{@{}lll@{}}
\textbf{ID} & \textbf{Description} & \textbf{Example Hashtags} \\ \toprule
1 & TV Shows & \textit{skamitalia, masterchefgr, thearchers, ibes} \\
10 & Fandoms & \textit{jungkook, choicefandom, saveshadowhunters} \\
19 & Art & \textit{painting, etsy, vintage, fasion, architecture, arte} \\
24 & Romance TV & \textit{loveisland, poweroflovegr, liebesgschichten} \\
44 & TV Promo & \textit{comingsoon, luciferonnetflix, skyupnext} \\
3 & Job Promo & \textit{career, hiring, jobs, startup, sales, jobsearch} \\
16 & Giveaways & \textit{giveaway, freebiefriday, sorteo, winwin, free} \\
31 & Music Promo & \textit{radio, youtube, hits, newmusic, magicfm, live} \\
11 & Government, News & \textit{bbcnews, afd, noafd, labour, parlament, orban} \\
23 & Covid, Crisis, Tech & \textit{covid19france, koronawirus, polizei, tech, gdpr} \\
30 & International Politics & \textit{france, syria, venezuela, eeuu, isis, migranti} \\
41 & Health & \textit{mentalhealth, autism, clapfornhs, discapacidad} \\
46 & Requests & \textit{stop, shoplocal, stopgiletsjaunes, helpme} \\
47 & Equality & \textit{metoo, 8demarzo, weltfrauentag, racisme, lgbt} \\
48 & Sports & \textit{arsenal, halamadrid, futbol, fcporto, rusia2018}
\end{tabular}}
\end{table}

\subsection{Measuring multilinguals' role across topics}

\subsubsection{Identifying topic-specific hashtags}

We train a multilingual contextualized topic model (CTM) to identify topics \citep{bianchi2021cross}. This approach uses multilingual sentence-BERT \citep{reimers2019sentence} as input to a variational autoencoder topic model to support zero-shot topic prediction for texts in unseen languages during training \cite{srivastava2017autoencoding,bianchi2021cross}. Crucially, the CTM uses the same topics in all languages, making direct comparisons across languages possible. The CTM is trained for 20 epochs on a random sample of 1M English-language tweets from the European decahose data. We set the total number of topics to 50 and retain default values for all other hyperparameters. We use the trained CTM to predict topic distributions for all tweets containing any hashtag highly associated with any language in any time interval. Tweets are assigned to the single topic with the highest probability, and hashtags are then assigned to single topics based on the most frequent topic of tweets in which it appears. 

We then manually inspect the ten most-frequent hashtags per topic to identify 15 topics of interest, where hashtags are coherent, meaningful, and reflect different types of content. These 15 topics account for 52.1\% of European tweets from our dataset, and 66.8\% of unique hashtags associated with any language at any time, which rises to 82.5\% when accounting for hashtag frequency (see topic distributions in Figure \ref{fig:topic_dist} in Supplemental Material). Brief topic descriptions are in Table \ref{tab:topic_examples}. To understand how content shapes multilinguals'  influence, we separate the 15 topics into four macro-categories: entertainment, politics, sports, and promotion (e.g., giveaways). The distribution of hashtag macro-categories across languages are shown in Figure \ref{fig:lang_hashtag} in the Supplemental Material.

\subsubsection{Evaluation}

We evaluated our multilingual hashtag topic assignment method with intrusion tests, following common practice \cite{hoyle2021automated}. Four hashtags were sampled from each topic $t$ with frequency weighting, and one hashtag was sampled from one of the 49 other topics based on frequency. Annotators tried to identify which hashtag does \textit{not} belong to $t$. We evaluated the 15 topics of interest in English, German, Spanish, Italian, and Turkish, selected based on annotators' language proficiencies. We conducted ten intruder tests for each topic in English, and 5 for each topic in the other languages. Three annotators completed the intruder test for English hashtags with an average accuracy of 0.73 and interannotator agreement of 0.67 (Krippendorff $\alpha$). Among tasks where at least two annotators selected the same intruder, accuracy rose to 0.78, and further rose to 0.89 for tasks where all three annotators agreed.  

\begin{table}[!htbp]
\caption{Annotator accuracy on topic intrusion tests, averaged over 3 annotators for English.}
\resizebox{\columnwidth}{!}{
\begin{tabular}{@{}cccccc@{}}
Language & English & Spanish & German & Turkish & Italian \\ \toprule
Accuracy & 0.731 & 0.747 & 0.813 & 0.587 & 0.627 \\
\end{tabular}
}

\label{tab:topic_eval}
\end{table}

\subsubsection{Estimating causal effects}

We extend Study 3's design to compare multilinguals' effects across topics. For a given MCP $(C_x,C_y)$ and locus $C_x$, units are $L_x$ monolinguals from $C_x$ and the treatment is having at least one multilingual $(L_x,L_y)$ contact. For each topic $t$, the outcome is whether a user shares a hashtag that belongs to topic $t$ and is associated with $L_y$. Our analysis focuses on overall causal effect estimates for all 15 topics. We also estimate locus-specific effects for each macro-category, with a summary of results and maps in the Supplemental Material (Figure \ref{fig:hashtag_maps}).

\begin{figure}[th]
    \centering
    \includegraphics[width=\linewidth]{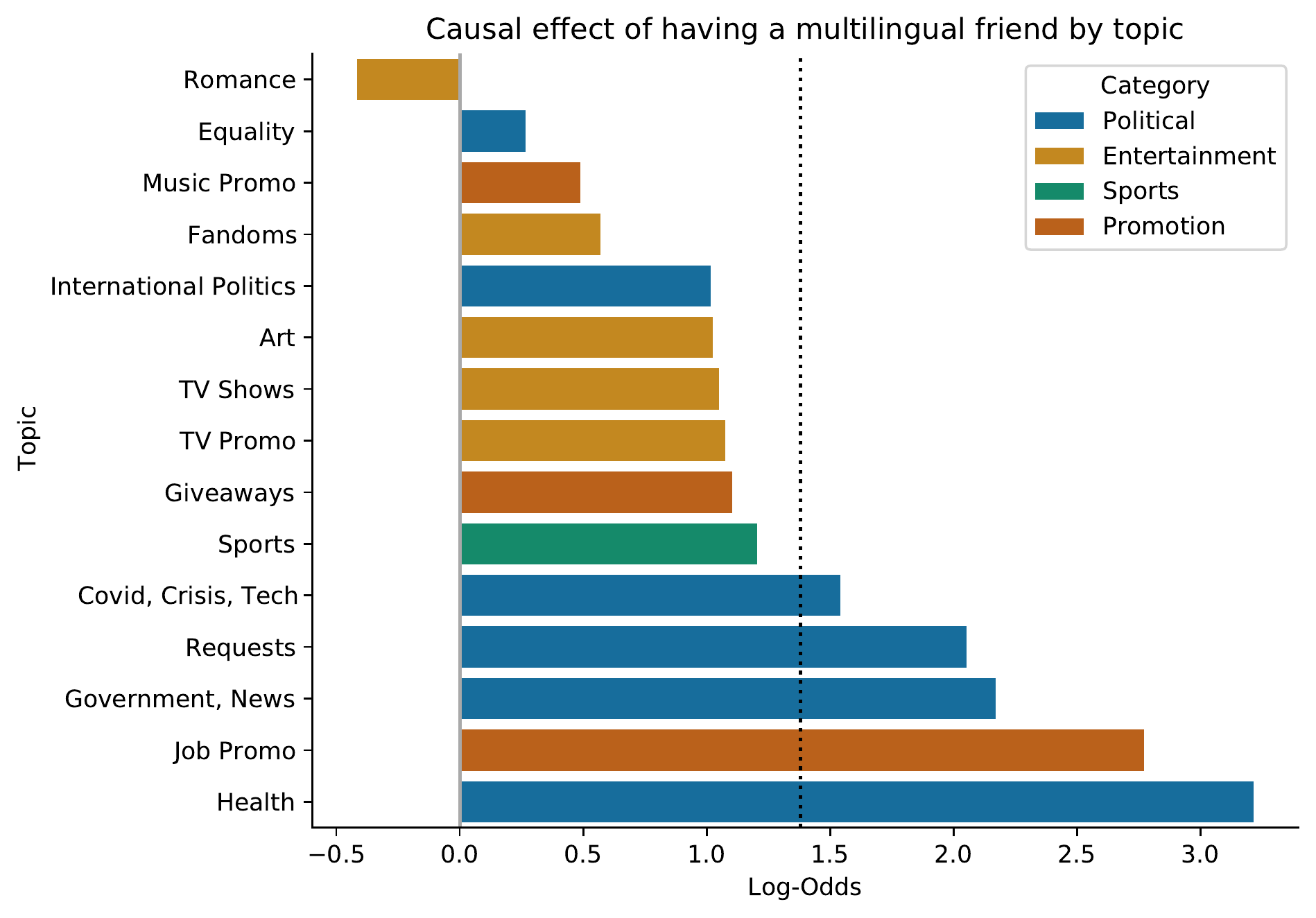}
    \caption{Causal effect (ATT) of having a multilingual friend on cross-lingual hashtag sharing across topics. All estimates have heteroskedasticity-consistent standard errors below 0.03 and are significant (p$<$0.0001). The x-axis shows log odds ratios, with 0 being no effect of having a multilingual friend, and the dashed line at 1.38 shows the overall effect on hashtag sharing. Colors represent macro-categories.}
    \label{fig:topic_effects}
\end{figure}

\subsection{Results}
Figure \ref{fig:topic_effects} shows the effect of multilinguals by hashtag topic and macro-category. Three key findings support our hypothesis that multilinguals play a greater role in spreading information that is otherwise less accessible to their peers. 

First, multilinguals have a greater communication influence on the cross-lingual diffusion of political content than entertainment. Political discourse likely occurs more in regional or country-centric public spheres \citep{schunemann2020ready}, so there is a greater reliance on multilingual individuals to broker political information across borders. Correspondingly, of all political topics, multilinguals play a smaller role for topics with widespread transnational popularity and awareness, such as: \textit{Equality}, which includes hashtags for global gender equality movements like \#metoo and \#8m, and \textit{International Politics}, which includes country names and non-European political organizations. Contrary to most political hashtags, entertainment hashtags often reflect globally-popular phenomena (e.g., K-pop fandoms) despite being associated with specific languages, and more cross-lingual information cascades involve entertainment content \citep{jin2017detection}. We believe that individual multilingual contacts play a smaller role in entertainment because there are more ways for that content to spread.   

Second, we argue that multilingual individuals are crucial for nascent social movements to gain global traction, but have a relatively small influence within well-established transnational movements. Figure \ref{fig:topic_effects} indicates that multilinguals have a large impact on spreading health-related information, which includes hashtags advocating for mental health awareness, COVID-related activism, and disability rights. In contrast to racial and gender equality movements reflected in the \textit{Equality} topic, organizing for disability rights gained traction as a social movement later than for race and gender both offline and on Twitter, where the public sphere about disability is still growing \citep{scotch1988disability,sarkar2021good}. The initial adoption of burgeoning social movements across countries relies heavily on direct contacts, such as multilingual friends \citep{mcadam1993cross}. Information diffusion about more well-established social movements do not depend on multilingual bridges because it occurs via many channels of communication, including news and television media \citep{mcadam1993cross}.

Third, multilinguals play an especially important role in sharing information about job searches and career opportunities. This parallels \citet{granovetter1973strength}'s argument about the strength of weak ties for job-seeking purposes. Even though monolingual users' ties with multilinguals are not necessarily weak, multilinguals similarly serve as bridges between different parts of a social network and thus facilitate access to novel information, such as job opportunities that users may not have otherwise been aware of.

\section{Discussion}

Gaining a complete picture of global information diffusion requires understanding how information crosses languages. We design three studies to investigate how multilingual users participate in this process, which we use to quantify their structural role and communication influence in information exchange across European languages on Twitter. For each pair of countries with different dominant languages, we construct networks where users are connected if they have mutually ``mentioned" each other. We use these networks in Study 1 to quantify the extent to which multilinguals' positions in these networks facilitate spreading novel information, as measured by betweenness centrality. In Studies 2 and 3, we quantify how having a multilingual contact influences the odds of monolingual users posting content from the other country's dominant language.

Results from all three studies show that multilinguals play an outsize role in cross-lingual information exchange compared to their monolingual peers. Effects vary widely but systematically across country pairs and topics. We conduct regression analyses to measure how the role of multilinguals is associated with demographic, geographic, economic, political, and linguistic aspects of the relationship between country pairs. To compare multilinguals' influence across topics, we augment our study design for hashtag sharing with multilingual contextualized topic modeling. 

In general, multilingual individuals have a greater influence on the spread of information that is otherwise \textit{less} accessible to their monolingual peers, as they play more of a gatekeeping role. Multilinguals have a greater effect on information diffusion between dominant languages of countries that are geographically far apart, with Western European multilinguals who post in Eastern European languages having an especially big influence. We identify a similar pattern for topics, where multilinguals have greater influence on cross-lingual information exchange for topics discussed in more restricted public spheres: national or regionally-oriented politics over entertainment which can have international appeal, nascent health-related social movements over established racial and gender equality movements, and job opportunities previously known only to small communities.

We acknowledge that this work has important limitations. First, our studies do not account for multilinguals who use minority languages (e.g., Basque) or reside in highly multilingual countries (e.g., Switzerland). Imperfect performance of location inference and language detection also limited the set of countries and languages studied. Furthermore, we make the simplifying assumption that tweets are written in one language, which does not adequately account for code-mixing within posts and the users who engage in such practices \citep{androutsopoulos2015networked}. While code-mixed tweets are a relatively small percent of Twitter communication \citep{rijhwani2017estimating}, accurately recognizing these tweets at scale has proven challenging due to the absence of labeled data for training models \citep{jurgens2017incorporating}. As the performance and efficiency of language detection of code-mixed tweets improve, we anticipate that incorporating such information would be fruitful, and analyzing the relationship between code-mixing strategies and information diffusion could yield interesting theoretical insights.

To avoid making assumptions about people's offline language usage or competence, we intentionally define multilinguals based on their performance on Twitter. However, this presents a significant limitation: users who only tweet in one language but understand multiple may also play an important, and perhaps different, role in information diffusion. Future research could employ different methodologies to highlight these users, such as linking social media activity with survey data about users' language backgrounds.

Further research can also improve upon our study designs. We adopt a traditional causal inference setup, which considers treatment status binary to emulate randomized experiments. Thus, all of our studies involve collapsing underlying continuous variables into binary indicators. A possible next step would involve adapting our studies to account for continuous treatments; this would facilitate investigation of how cross-lingual information exchange is impacted by a user's degree of multilingualism (Study 1) or the number and/or strength of a user's ties to multilinguals (Studies 2 and 3).  

Beyond addressing these limitations, there are numerous avenues for future work. For example, we adopt a microscopic perspective on information diffusion by examining how multilingualism impacts individual users' roles in information diffusion; we focus on local influence because it is more precise and less random than observations of information cascades \citep{bakshy2011everyone}. Nevertheless, an interesting extension that takes a macroscopic perspective, perhaps involving simulations of cross-lingual information cascades \citep{chen2021burden}, could help contextualize how these individual-level effects, in aggregate, shape the global flow of information. Another future direction would involve considering other forms of information that may spread via different mechanisms than URLs and hashtags, such as meme templates, images, videos, and text outside of hashtags. There is likely variation in the role of multilinguals across semantic dimensions beyond topic, such as emotional valence or misinformation. Finally, future research can assess the generalizability of our findings beyond the scope of European Twitter by applying our methodology to study other regions, languages, and platforms.

\section{Broader impact and ethical considerations}

Understanding the role of multilinguals in information diffusion has immense consequences. Platforms like Twitter can empower multilinguals to spread information that supports positive outcomes such as knowledge-sharing, collaboration, crisis response, or social progress, thus enabling different language communities to benefit from a truly global social network \citep{eleta2014multilingual,hale2014global}. Our study not only highlights this potential but also identifies how it varies across topics and with respect to the geographical, linguistic, and political relationship between the countries. For instance, our research suggests that multilinguals can be better utilized to spread political news as opposed to entertainment. We also see that their importance is more pronounced for supporting information spread across countries further away from each other. Such findings not only highlight the contexts where multilinguals already play an important role but also help us identify the barriers for cross-lingual diffusion; in such situations, platforms may benefit more from technologies such as machine translation.

Our research can also help platforms address dangerous consequences of global networks by focusing efforts on nudging multilinguals to mitigate the spread of harmful information such as misinformation, conspiracy theories, or online abuse. Past network science research shows the value of betweenness centrality in identifying nodes that can limit the spread of such information \citep{golovchenko2018state}. Here, we show that multilingual nodes tend to have high betweenness centrality. Furthermore, our study shows that multilinguals play a particularly important role in the spread of political topics, common targets for malicious actors aiming to spread propaganda and disinformation.

Despite the potential for positive impact, we acknowledge the ethical risks of this work. Rather than stem the flow of harmful content, our work may inspire malicious agents to target and manipulate multilinguals into propagating such information. In addition, our focus on users of politically and socially-dominant language varieties and use of automated language detection excludes people whose posts contain endangered or minority languages, non-prestige dialects of dominant languages, or code-mixing. Although our work does not present direct harm to individuals, these decisions systematically exclude marginalized groups whose online behavior deserves equal consideration.

To promote transparency and future research, we publicly share data, code, and models but take steps to preserve user privacy. The datasheets shared for causal effect estimation include only variables necessary to replicate our results. We do not share user IDs, raw text, or other personally-identifiable information. While the location inference tool used presents a privacy risk by inferring users' specific geo-coordinates, we only store information at the country level.

\section{Conclusion}
By developing a set of causal inference studies that measure users' structural role and communication influence, we show that multilingually-posting users on European Twitter are particularly important for information diffusion across languages. These users have an especially large influence in situations where they serve more as gatekeepers in information flow, particularly in spreading information from places and topics that are otherwise inaccessible to their monolingual peers. This work is crucial for understanding how information is shared around the world, and has implications for platforms to support beneficial consequences of global social networks while mitigating potential harms. Publicly-available code, models, and aggregated data can be found at: \\\url{https://github.com/juliamendelsohn/bridging-nations}.

\section*{Acknowledgments}
This research was supported by the National Science Foundation (Grants IIS-1815875 and IIS-2007251) and through funding from the Volkswagen Foundation.

\small{\bibliography{bibfile.bib}}

\appendix
\setcounter{table}{0}
\renewcommand{\thetable}{S\arabic{table}}
\setcounter{figure}{0}
\renewcommand{\thefigure}{S\arabic{figure}}

\onecolumn
\section*{Supplemental Material}
\normalsize

\begin{table}[htbp!]
\caption{All countries initially considered, the set of included countries with dominant language code, and set of excluded countries after filtering.}
\label{tab:countries}
\resizebox{\textwidth}{!}{%
\begin{tabular}{|lll|lll|ll|} \hline
\multicolumn{6}{|c|}{Included Countries} & \multicolumn{2}{c|}{Excluded Countries} \\ \hline
Alpha-2 & Country & Language & Alpha-2 & Country & Language & Country & Reason \\ \hline
AM & Armenia & hy & IE & Ireland & en & Andorra & size \\
AT & Austria & de & IL & Israel & iw & Albania & no Albanian tweets \\
AZ & Azerbaijan & tr & IS & Iceland & is & Bosnia & multilingual \\
BG & Bulgaria & bg & IT & Italy & it & Belgium & multilingual \\
BY & Belarus & ru & LT & Lithuania & lt & Switzerland & multilingual \\
CZ & Czechia & cs & LV & Latvia & lv & Cyprus & multilingual \\
DE & Germany & de & MD & Moldova & ro & Croatia & 67\% undetected lang \\
DK & Denmark & da & NL & Netherlands & nl & Liechtenstein & size \\
EE & Estonia & et & NO & Norway & no & Luxembourg & multilingual \\
ES & Spain & es & PL & Poland & pl & Monaco & size\\
FI & Finland & fi & PT & Portugal & pt & Montenegro & multilingual\\
FR & France & fr & RO & Romania & ro & North Macedonia & multilingual \\
GB & United Kingdom & en & RU & Russia & ru & Malta & size \\
GE & Georgia & ka & SE & Sweden & sv & Serbia & 70\% undetected lang \\
GR & Greece & el & SI & Slovenia & sl & Slovakia & no Slovak tweets \\
HU & Hungary & hu  & TR & Turkey & tr & San Marino & size  \\
 &  &  &  &  &  & Ukraine & multilingual \\
 &  &  &  &  &  & Vatican City & size \\ \hline

\end{tabular}%
}
\end{table}

\begin{table}[htbp!]
\centering
\caption{All eligible multilingual country pairs (MCPs). Abbreviations are ISO 3166-1 alpha-2 country codes.}
\label{tab:mcp}
\resizebox{.75\textwidth}{!}{%
\begin{tabular}{|llllllll|} \hline
(AM, IE) & (AZ, SE) & (DK, IT) & (FI, SE) & (DE, NL) & (IE, IL) & (LT, LV) & (PL, PT) \\
(AM, RU) & (AZ, GB) & (DK, NL) & (FI, TR) & (DE, NO) & (IE, IT) & (LV, NL) & (PL, RU) \\
(AM, GB) & (BG, BY) & (DK, NO) & (FI, GB) & (DE, PL) & (IE, LV) & (LV, PL) & (PL, SI) \\
(AT, CZ) & (BY, CZ) & (DK, PL) & (DE, FR) & (DE, PT) & (IE, LT) & (LV, RU) & (ES, PL) \\
(AT, FI) & (BY, FR) & (DK, ES) & (FR, GR) & (DE, RU) & (IE, NL) & (ES, LV) & (PL, SE) \\
(AT, FR) & (BY, DE) & (DK, SE) & (FR, HU) & (DE, ES) & (IE, NO) & (LV, TR) & (PL, TR) \\
(AT, GR) & (BY, IE) & (DK, TR) & (FR, IS) & (DE, SE) & (IE, PL) & (GB, LV) & (GB, PL) \\
(AT, IE) & (BY, LV) & (DK, GB) & (FR, IE) & (DE, TR) & (IE, PT) & (LT, NL) & (PT, RO) \\
(AT, IT) & (BY, PL) & (EE, FI) & (FR, IL) & (DE, GB) & (IE, RO) & (LT, PL) & (PT, RU) \\
(AT, NL) & (BY, ES) & (EE, FR) & (FR, IT) & (GR, IE) & (IE, RU) & (LT, PT) & (ES, PT) \\
(AT, NO) & (BY, TR) & (DE, EE) & (FR, LV) & (GR, IT) & (IE, SI) & (ES, LT) & (PT, SE) \\
(AT, PL) & (BY, GB) & (EE, IE) & (FR, LT) & (ES, GR) & (ES, IE) & (LT, SE) & (PT, TR) \\
(AT, PT) & (BG, IE) & (EE, IT) & (FR, NL) & (GR, TR) & (IE, SE) & (LT, TR) & (GB, PT) \\
(AT, RU) & (BG, RU) & (EE, NL) & (FR, NO) & (GB, GR) & (IE, TR) & (GB, LT) & (ES, RO) \\
(AT, ES) & (BG, GB) & (EE, PL) & (FR, PL) & (HU, IE) & (IL, RU) & (NL, NO) & (RO, TR) \\
(AT, SE) & (CZ, FR) & (EE, PT) & (FR, PT) & (HU, IT) & (ES, IL) & (NL, PL) & (GB, RO) \\
(AT, TR) & (CZ, DE) & (EE, RU) & (FR, RO) & (HU, NL) & (GB, IL) & (NL, PT) & (ES, RU) \\
(AT, GB) & (CZ, IE) & (EE, ES) & (FR, RU) & (HU, PL) & (IT, LV) & (NL, RO) & (RU, SE) \\
(AZ, BY) & (CZ, IT) & (EE, SE) & (FR, SI) & (HU, PT) & (IT, LT) & (NL, RU) & (RU, TR) \\
(AZ, FI) & (CZ, PL) & (EE, TR) & (ES, FR) & (ES, HU) & (IT, NL) & (ES, NL) & (GB, RU) \\
(AZ, FR) & (CZ, RU) & (EE, GB) & (FR, SE) & (HU, SE) & (IT, PL) & (NL, SE) & (ES, SI) \\
(AZ, DE) & (CZ, SI) & (FI, FR) & (FR, TR) & (HU, TR) & (IT, PT) & (NL, TR) & (SI, TR) \\
(AZ, IE) & (CZ, ES) & (DE, FI) & (FR, GB) & (GB, HU) & (IT, RO) & (GB, NL) & (GB, SI) \\
(AZ, IT) & (CZ, SE) & (FI, IE) & (DE, GR) & (IE, IS) & (IT, RU) & (NO, PL) & (ES, SE) \\
(AZ, NL) & (CZ, TR) & (FI, IT) & (DE, HU) & (IS, NL) & (IT, SI) & (NO, PT) & (ES, TR) \\
(AZ, PL) & (CZ, GB) & (FI, NL) & (DE, IE) & (ES, IS) & (ES, IT) & (ES, NO) & (ES, GB) \\
(AZ, PT) & (DK, FR) & (FI, PT) & (DE, IT) & (IS, SE) & (IT, SE) & (NO, SE) & (SE, TR) \\
(AZ, RU) & (DE, DK) & (FI, RU) & (DE, LV) & (IS, TR) & (IT, TR) & (NO, TR) & (GB, SE) \\
(AZ, ES) & (DK, IE) & (ES, FI) & (DE, LT) & (GB, IS) & (GB, IT) & (GB, NO) & (GB, TR) \\ \hline
\end{tabular}%
}
\end{table}

\begin{table}[htp!]
\centering
\caption{Intercoder agreement (Krippendorff's $\alpha$) between each pair of LangID models. Bolded values represent the highest agreement score for each column. Twitter's LangID model has the highest average agreement with other models.}
\begin{tabular}{ccccccc}
\toprule
Model & Twitter & fastText & langid.py & langdetect & CLD2 & CLD3 \\ \midrule
Twitter &  - & \textbf{0.87} & 0.84 & \textbf{0.82} & \textbf{0.82} & \textbf{0.77} \\
fastText & \textbf{0.87} &  - & \textbf{0.86} & 0.80 &  0.79 & 0.75 \\
langid.py & 0.84 & 0.86 &  - & 0.79 & 0.76 & 0.74 \\
langdetect & 0.82 & 0.80 & 0.79 & - & 0.70 & 0.70 \\
CLD2 & 0.82 & 0.79 & 0.76 & 0.70 &  - & 0.71 \\
CLD3 & 0.77 & 0.75 & 0.74 & 0.70 & 0.71 & - \\ \midrule
Mean & 0.824 & 0.814 & 0.798 & 0.763 & 0.760 & 0.734 \\ \bottomrule
\end{tabular}
\label{tab:langid}
\end{table}

\begin{table}[htp!]
\centering
\caption{Examples of hashtags (from one selected time interval) and domains associated with five different languages. }
\label{tab:examples}
\begin{tabular}{@{}ccccc@{}}
\toprule
German & Portuguese & Turkish & Polish & English \\ \midrule
cdu & fcporto & çağlarertuğrul & pis & oddoneout \\
spd & todosportugal & sustunuz & konwencjapis & remain \\
merkel & capricórnio & pazartesi & topmodel & eastenders \\
klimaschutz & aquário & cumartesi & thevoiceofpoland & liarjohnson \\
noafd & sportingcp & burcuözberk & kaczyński & ncfc \\ \hline \hline
tagesschau.de & publico.pt & tele1.com.tr & wpolityce.pl & manchestereveningnews.co.uk \\
faz.net & record.pt & haber.sol.org.tr & niezalezna.pl & whounfollowedme.org \\
spiegel.de & maisfutebol.iol.pt & diken.com.tr & dorzeczy.pl & theneweuropean.co.uk \\ \bottomrule
\end{tabular}
\end{table}

\begin{figure*}[htp!]
    \includegraphics[width=\linewidth]{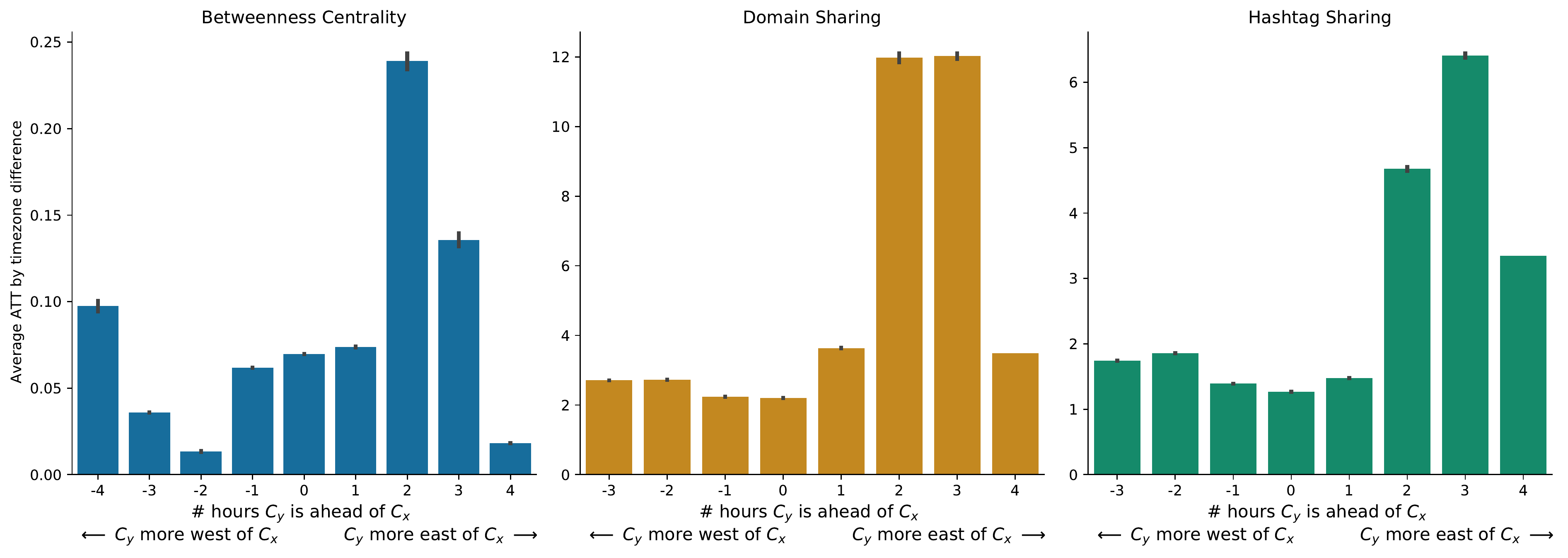}
    \caption{Treatment effect of multilinguals on bringing information associated with country $C_y$'s dominant language to country $C_x$ as a function of the timezone difference between the two countries. To account for population differences,  we average ATTs over all treated users, even though there is a single ATT value per country in each pair. The relationship between ATT and timezone is shown for all three measures of influence: betweenness centrality (Study 1, left), domain sharing (Study 2, center), and hashtag sharing (Study 3, right). Effects are greatest when $C_y$ is 2-3 hours ahead (i.e. much further east) of $C_x$.}
    \label{fig:time_diff}
\end{figure*}

\begin{figure*}[htp!]
    \includegraphics[width=\linewidth]{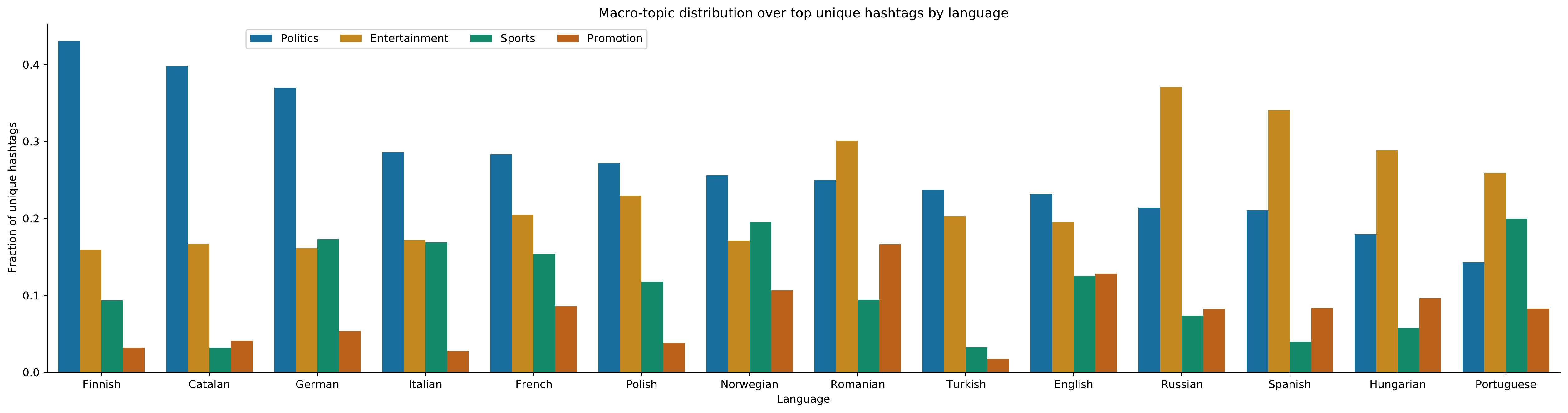}
    \caption{Fraction of unique hashtags associated with sample of languages for each of four macro-topics, sorted left to right from largest to smallest proportion of political hashtags. Finnish and German hashtags are largely political, while Russian, Spanish, and Hungarian hashtags tend to be more entertainment-oriented. Portuguese hashtags are much more focused on entertainment and sports compared to politics. This figure also illustrates how language choice can be a meaningful sociolinguistic variable. For example, while Spanish hashtags are more entertainment-oriented, Catalan hashtags tend to be more political; this is consistent with prior work that shows Catalan is used online to mark political identity \cite{stewart2018s}.}
    \label{fig:lang_hashtag}
\end{figure*}

\begin{figure*}[htp!]
    \includegraphics[width=\linewidth]{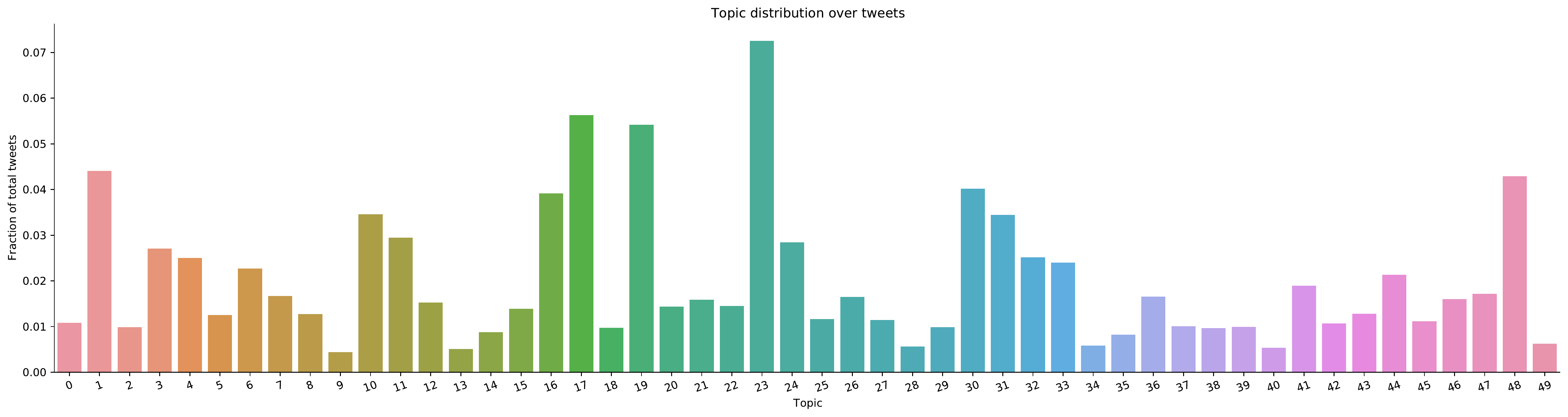}
    \includegraphics[width=\linewidth]{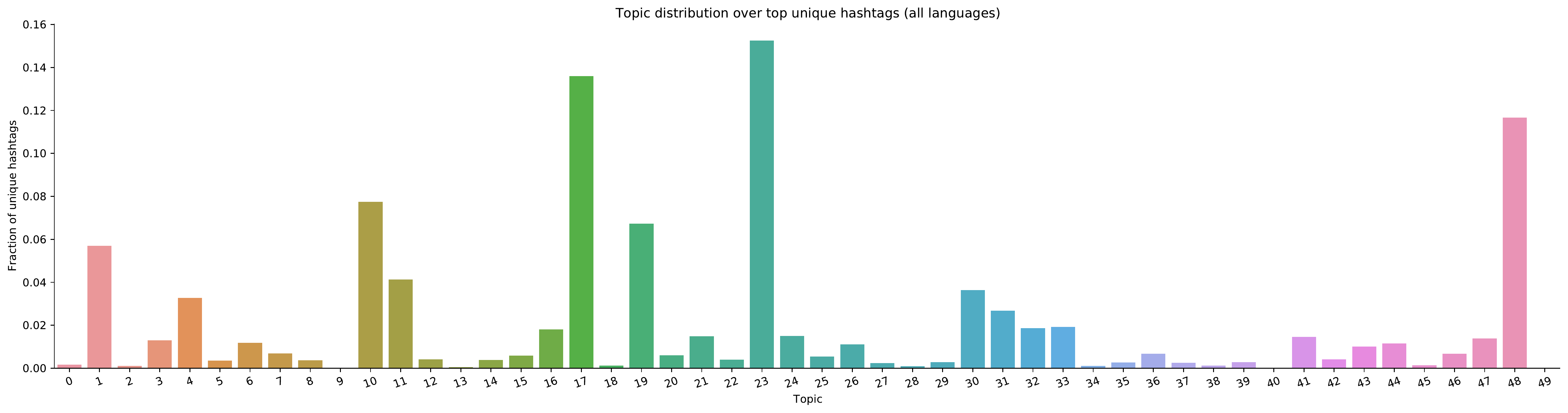}
    \includegraphics[width=\linewidth]{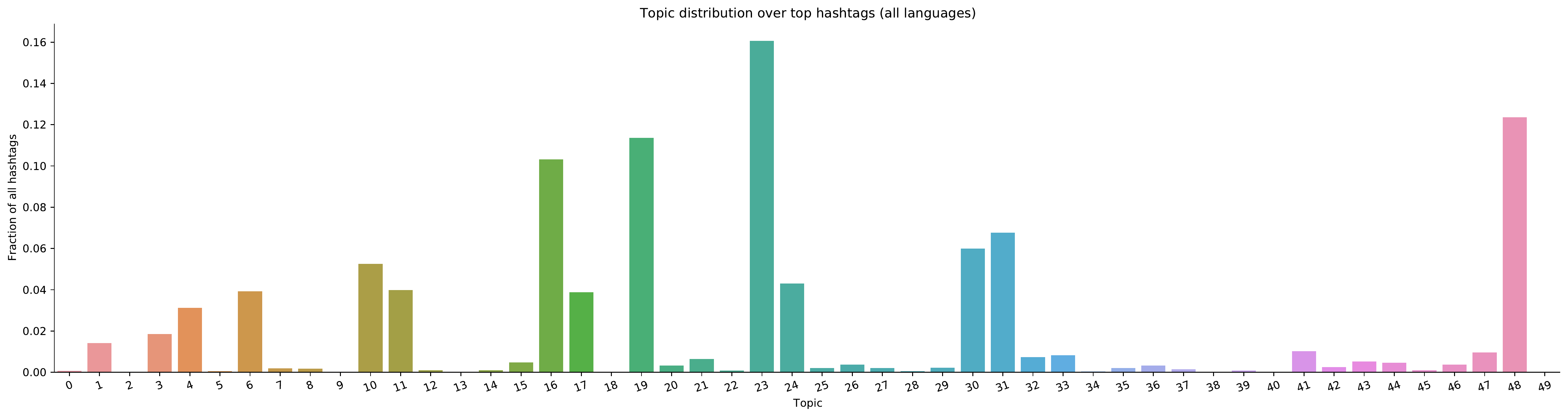}
    \caption{Distributions of tweets and hashtags assigned to all 50 topics with multilingual contextualized topic modeling. The top plot shows the fraction of all tweets assigned to each topic. The middle plot shows the fraction of unique hashtags assigned to each topic (out of all hashtags associated with at least one language in one time interval), and the bottom shows distribution of total hashtags over topics accounting for frequency in the Decahose data.}
    \label{fig:topic_dist}
\end{figure*}

\begin{table}[!htbp]
\centering
\caption{Summary of locus-specific causal estimates for each topic macro-category. Out of the 464 possible loci from 232 MCPs, ATT estimates were only calculated for a subset of loci with sufficient data. The remaining columns show the percent of estimates where multilinguals have a significantly positive, negative, and no significant effect (based on heteroskedasticity-consistent standard error estimation with $p < 0.05$.}
\label{tab:effect_proportions}
\begin{tabular}{@{}ccccc@{}}
\toprule
\textbf{Macro-Topic} & \begin{tabular}[c]{@{}c@{}}\textbf{Num.}\\ \textbf{Estimates}\end{tabular} & \textbf{\% Positive} & \textbf{\% Negative} & \begin{tabular}[c]{@{}c@{}}\textbf{\% No Sig.}\\ \textbf{Effect}\end{tabular} \\ \midrule
Politics & 240 & 0.433 & 0.054 & 0.513 \\ 
Entertainment & 228 & 0.474 & 0.035 & 0.491 \\ 
Sports & 199 & 0.432 & 0.030 & 0.538 \\ 
Promotion & 146 & 0.370 & 0.034 & 0.596 \\ \bottomrule
\end{tabular}
\end{table} 

\begin{figure*}[ht]
  \centering
    \includegraphics[width=.5\linewidth]{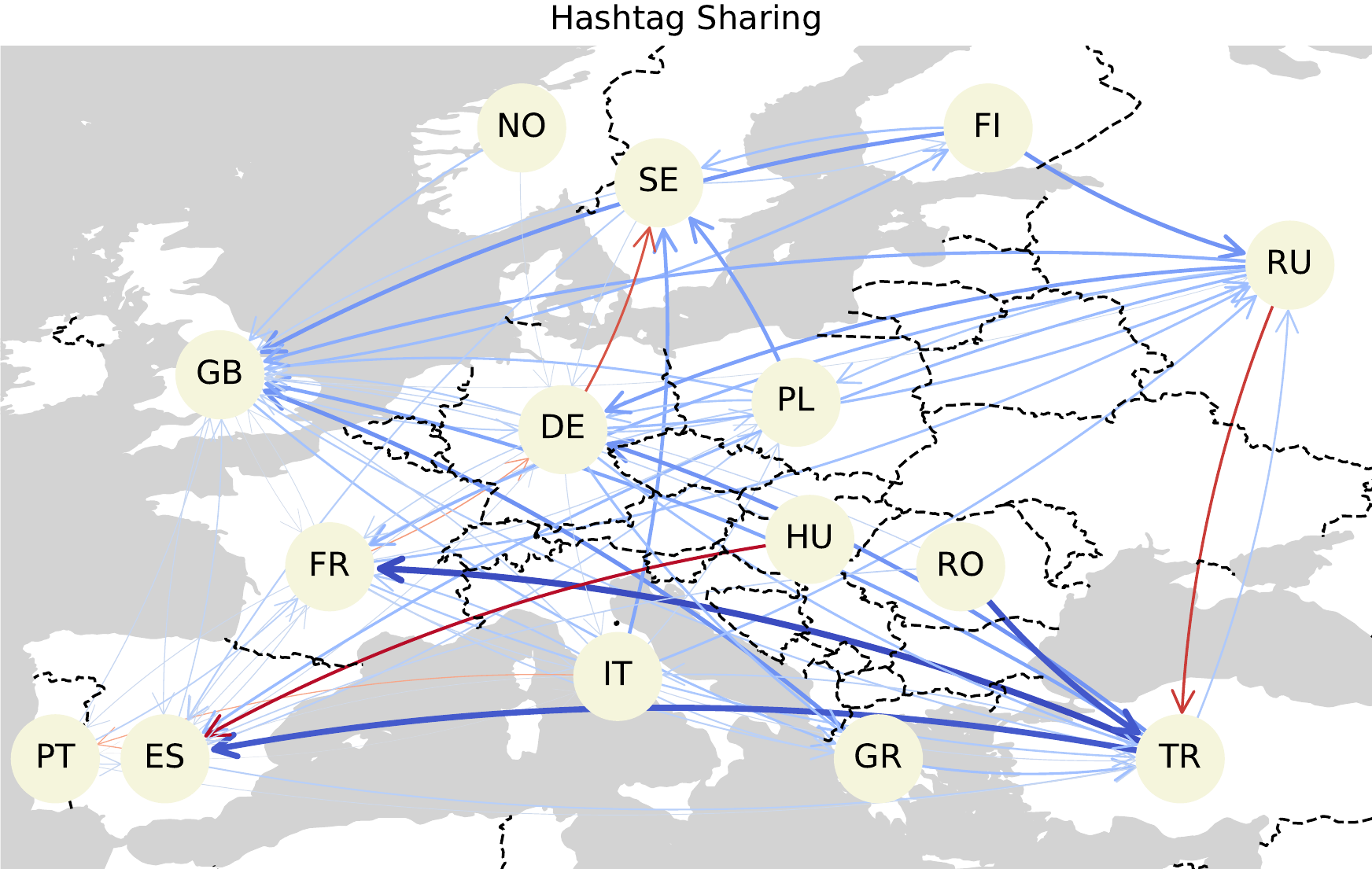}\hfill
    \includegraphics[width=.5\linewidth]{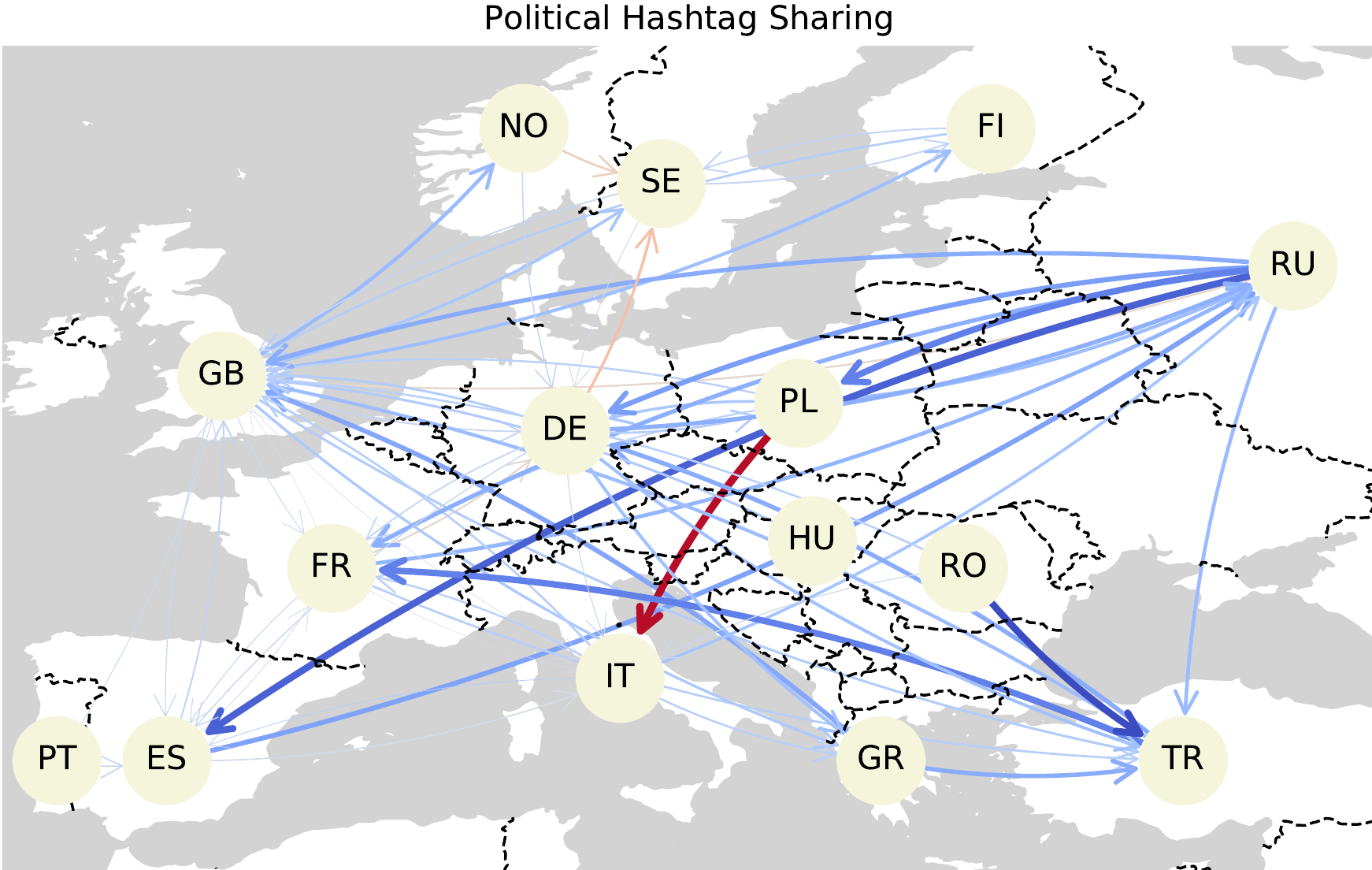}\hfill
    \includegraphics[width=.5\linewidth]{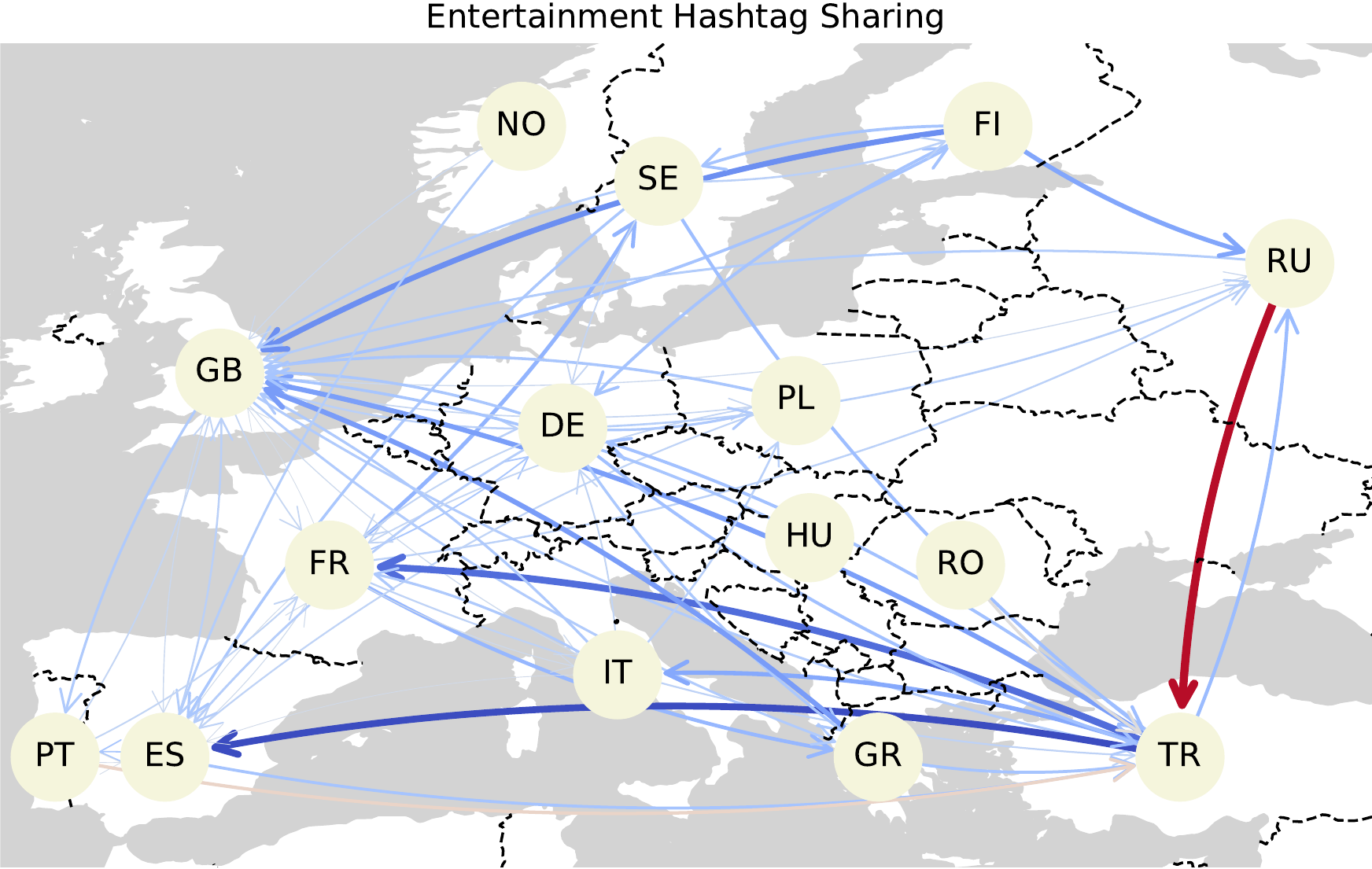}\hfill
    \includegraphics[width=.5\linewidth]{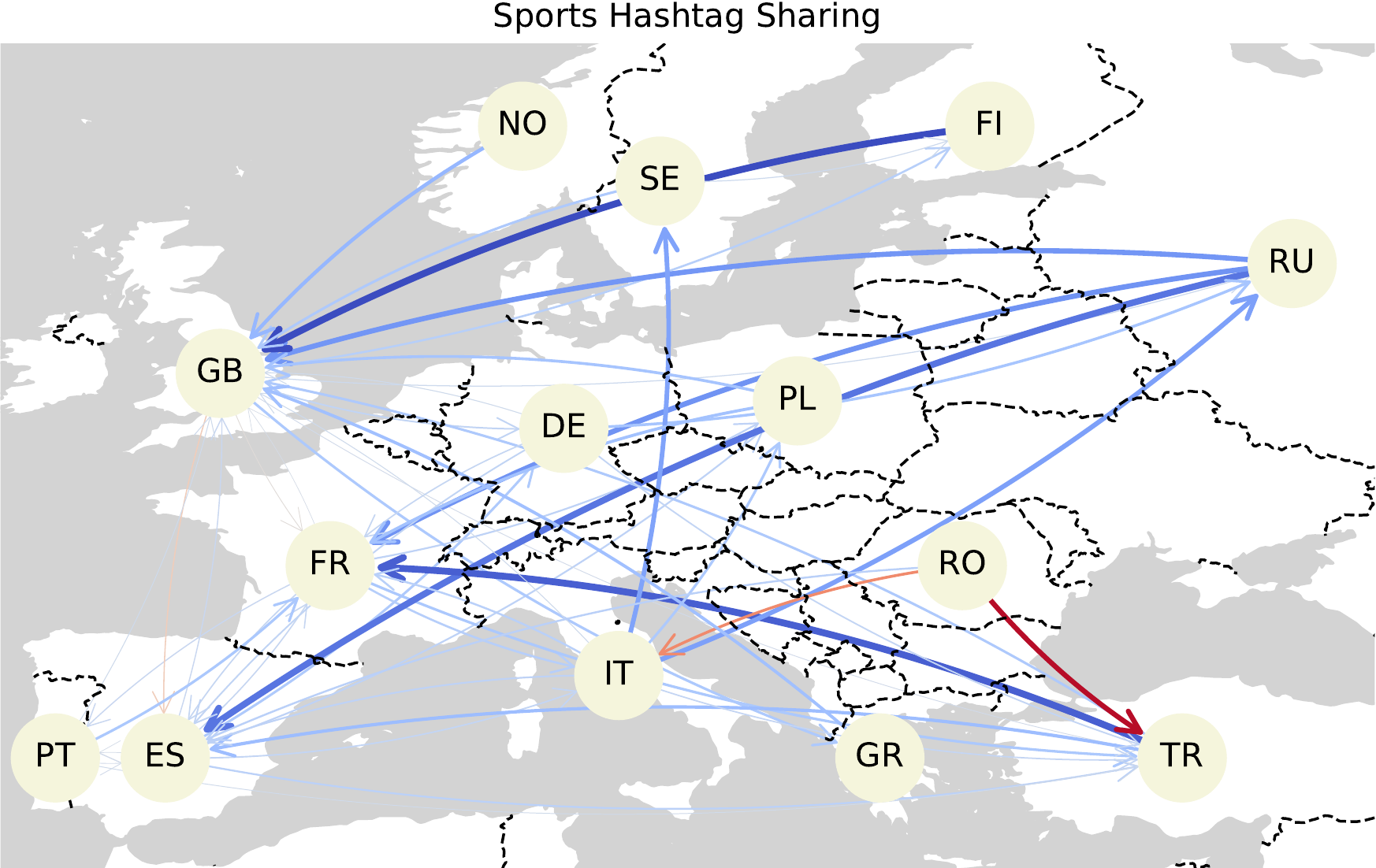}\hfill
    \includegraphics[width=.5\linewidth]{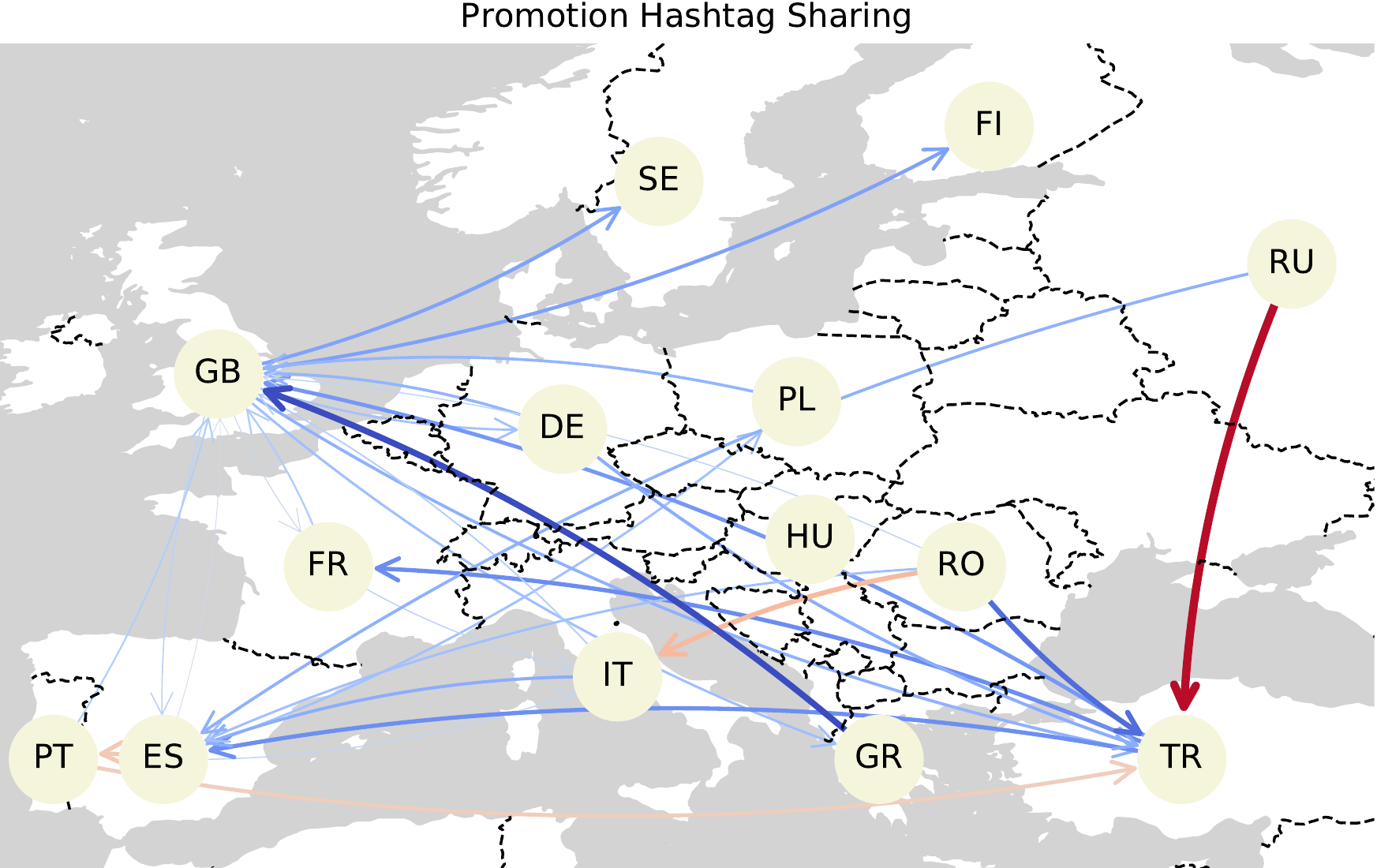}\hfill
    \includegraphics[width=.4\linewidth]{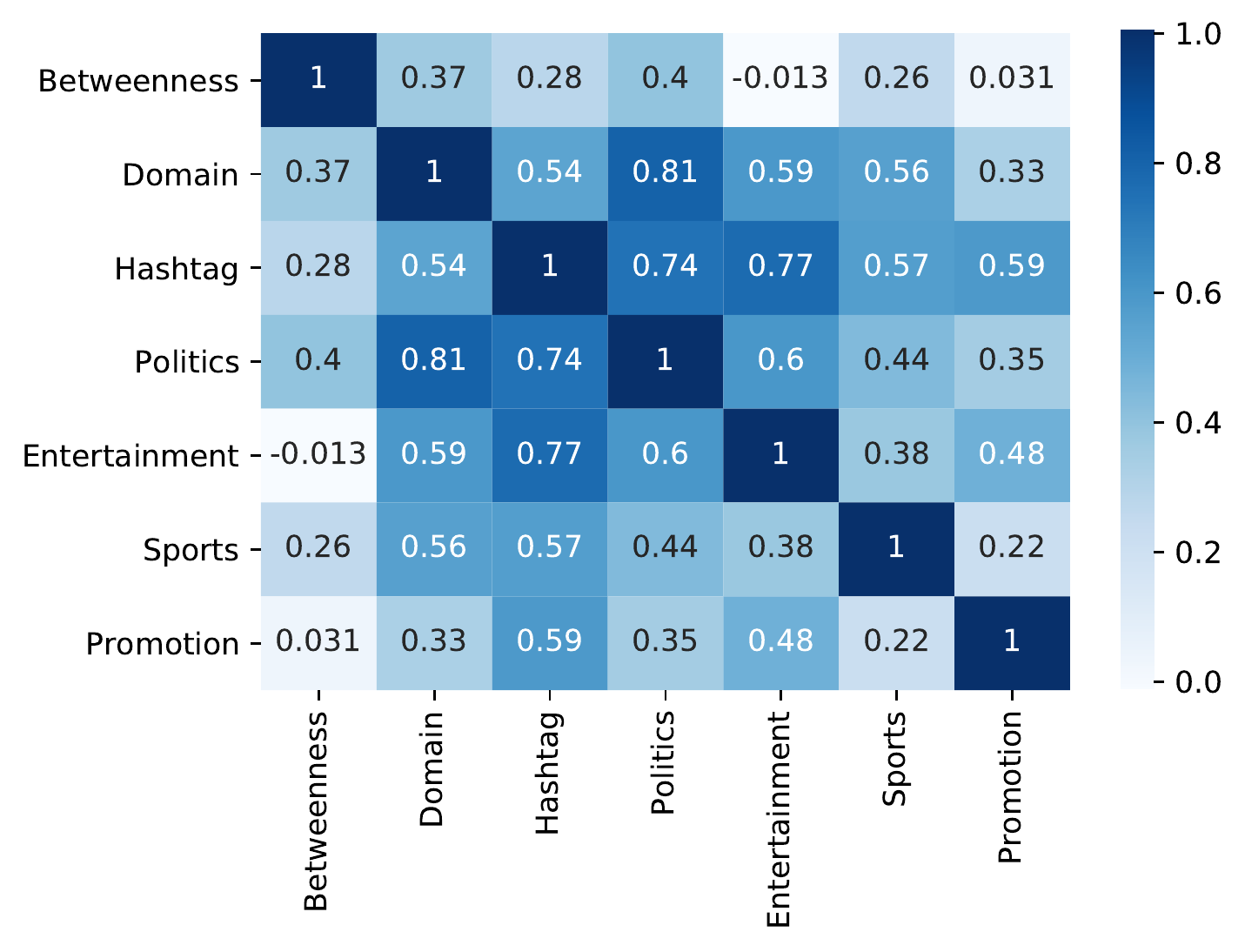} \hfill

    \caption{Causal effect (ATT) of having a multilingual friend on overall cross-lingual hashtag sharing (top left) and four macro-topics: politics (top right), entertainment (middle left), sports (middle right), and promotion (bottom right). Each edge indicates effect on the likelihood of monolingual users from the destination node to share a hashtag associated with the source node's dominant language. Negative effects are in red and positive are blue. The magnitude of the causal effect is indicated by arrow shading and width. Only statistically significant estimates ($p<0.05$ with robust standard error estimation) are shown. \\
    The heatmap on the bottom right shows Pearson correlations between locus-specific causal estimates over all 7 outcomes (the 3 main studies, and 4 topic-specific outcomes). Given that the betweenness centrality outcome is based on a different sample, treatment, and estimator than the communication influence measurements, it is not surprising that betweenness centrality has the weakest correlation with other outcomes. Nevertheless, there is still positive correlation with most other outcomes, particularly domain and political hashtag sharing. Correlation between the communication influence outcomes are much higher, especially for domain and political hashtag sharing. This suggests that there may be commonalities in the type of content, with language-specific domains primarily focusing on news and political content.}
    \label{fig:hashtag_maps}
\end{figure*}

\end{document}